\documentclass[10pt, a4paper]{article}
\usepackage{amssymb,amsmath,amsfonts}
\usepackage[dvipsnames]{xcolor}
\usepackage{graphicx}
\usepackage{longtable}
\usepackage{verbatim}
\usepackage{color}
\usepackage{mdframed}
\usepackage{soul}
\usepackage{mathrsfs}
\usepackage{amsfonts,amssymb,mathrsfs,amsmath,esint}
\usepackage{slashed, cancel}
\usepackage{framed}
\usepackage{mdframed}
\usepackage{simplewick} 
\allowdisplaybreaks
\usepackage[bottom]{footmisc}
\usepackage{geometry}
\usepackage{tcolorbox}
\usepackage{latexsym}
\usepackage{graphicx}
\graphicspath{{./Figures/}}
\usepackage[dvipsnames]{xcolor}
\usepackage{booktabs}
\usepackage{cite}
\usepackage{datetime}
\newdateformat{mydate}{\THEDAY{ }\monthname[\THEMONTH]{ }\THEYEAR}

\usepackage[toc]{appendix} 

\usepackage{xcolor}
\usepackage{enumerate}
\usepackage{framed}
\usepackage{mdframed}
\usepackage{amsmath}
\usepackage{amsfonts}
\usepackage{mathrsfs}
\usepackage{amssymb}
\usepackage{graphicx, rotating}
\usepackage{epsfig}
\usepackage{latexsym}
\usepackage{graphicx}
\usepackage{color}
\usepackage{amsmath,bm,amssymb}
\usepackage{slashed}
\usepackage{hyperref}
\usepackage[font=footnotesize,labelfont=]{caption}
\hypersetup{colorlinks=true, citecolor=bluscuro, linkcolor=black, urlcolor=bluscuro}
\definecolor{rossos}{cmyk}{0,1,1,0.55}
\definecolor{bluscuro}{rgb}{0.15, 0.2, .85}
\definecolor{bluchiaro}{cmyk}{1,.3,0.,0.1}
\usepackage[T1]{fontenc} 
\usepackage{subfig}

\usepackage{tcolorbox}
\newcommand{\GC}{\text{\tiny GC}}
\newcommand{\PP}{\text{\tiny P}}
\newcommand{\LA}[0]{L_\text{\tiny A}}
\newcommand{\LB}[0]{L_\text{\tiny B}}

\newcommand{\GW}{\text{\tiny GW}}
\newcommand{\nn}{\nonumber}

\def\0{\vec{0}}

\newcommand{\GI}{\text{\tiny GI}}

\newcommand{\TT}{\text{\tiny TT}}

\newcommand{\bea}{\begin{eqnarray}}
\newcommand{\eea}{\end{eqnarray}}
\def\beq{\begin{equation}}
\def\eeq{\end{equation}}

\def\d{{\rm d}}

\def\beqa{\begin{eqnarray}}

	\def\eeqa{\end{eqnarray}}
\def\lsim{\mathrel{\rlap{\lower4pt\hbox{\hskip0.5pt$\sim$}}
		\raisE_1pt\hbox{$<$}}}         
\def\gsim{\mathrel{\rlap{\lower4pt\hbox{\hskip0.5pt$\sim$}}
		\raisE_1pt\hbox{$>$}}}         

\def\d{{\rm d}}

\def\d{{\rm d}}

\newcommand{\Ic}{\mathcal{I}_c}
\newcommand{\Is}{\mathcal{I}_s}

\def\eeqa{\end{eqnarray}}
\numberwithin{equation}{section}

\def\bq{\begin{quote}}
\def\eq{\end{quote}}

 at 10truept
\newcommand{\arXiv}[2]{\href{http://arxiv.org/pdf/#1}{{\tt [#2/#1]}}}
\newcommand{\arXivold}[1]{\href{http://arxiv.org/pdf/#1}{{\tt [#1]}}}

\newcommand{\llp}{\left [}
\newcommand{\rrp}{\right ]}

\newcommand{\lp}{\left (}
\newcommand{\rp}{\right )}

\renewcommand{\P}{\mathcal{P}}

\newcommand{\be}{\begin{equation}\begin{aligned}}
\newcommand{\ee}{\end{aligned}\end{equation}}

\usepackage[normalem]{ulem}

\def\wt{\widetilde}

\def\C{{\cal{C}}}

\def\H{{\cal H}}
\def\cs2{c_{\rm{s}}^2}

\def\U0{{\bar U_0}}
\def\wt{\widetilde}

\newcommand\gami[1]{{\gamma_{{#1}}^{~i}}}

\newcommand\A[1]{{\phi_{{#1}}}}

\def\X{{\cal{X}}}

\def\XB{{{\cal{X}}_{\rm{B}}}}

\def\U{{\cal{U}}}

\def\syn{\text{\tiny TT}}

\begin{document}
\parskip=10pt
\baselineskip=18pt

{\footnotesize  }
\vspace{5mm}
\vspace{0.5cm}

\begin{center}

\def\thefootnote{\fnsymbol{footnote}}

\topskip=70pt

{ \Large
\bf On the Gauge Invariance of Cosmological Gravitational Waves 
}
\\[1.2cm]

{V. De Luca\textit{\small $^{1}$}, 
G. Franciolini\textit{\small $^{1}$}, 
A. Kehagias\textit{\small $^{2}$} and 
A. Riotto\textit{\small $^{1}$}} \\[0.6cm]

{\small \textit{$^1$ D\'epartement de Physique Th\'eorique and Centre for Astroparticle Physics (CAP), \\
Universit\'e de Gen\`eve, 24 quai E. Ansermet, CH-1211 Geneva, Switzerland}}

{\small \textit{$^2$ Physics Division, National Technical University of Athens 15780 Zografou Campus, Athens, Greece} }

\vspace{.2cm}

\end{center}

\vspace{.8cm}

\begin{center}
\textbf{Abstract}
\end{center}
\noindent
The issue of the gauge invariance of gravitational waves arises if they are produced in the early universe at second-order in perturbation theory.
We address it  by dividing the discussion  in three parts: the production of gravitational waves, their propagation in the real universe, and their measurement. 

\vspace{.5in}

\def\thefootnote{\arabic{footnote}}
\setcounter{footnote}{0}
\pagestyle{empty}


\newpage
\pagestyle{plain}
\setcounter{page}{1}

\section{Introduction}
\noindent
The recent discovery of Gravitational Waves (GWs) produced by the merging of two massive black holes \cite{ligo} has started the new era of GW astronomy \cite{rev}. Aside from the astrophysical ones, there may be  many other  sources of GWs produced in the early universe. One of them  has been extensively studied in the literature and is related to the production of Primordial Black Holes (PBHs) from large curvature perturbations generated during inflation \cite{lrr}. These PBHs are generated through a collapse process once a sizeable small-scale fluctuation re-enters the Hubble radius. These large scalar perturbations generated in this scenario unavoidably provide a second-order source  of primordial GWs 
\cite{Acquaviva:2002ud, Mollerach:2003nq} at horizon re-entry \cite{e,jap,altri8,Ando:2017veq,sm,kohri, sas, Bartolo:2018evs,Bartolo:2018rku,unal,b1,b2}. The very same source has been studied in Refs. \cite{ Ananda:2006af, Baumann:2007zm} to investigate the GWs produced by the large-scale scalar perturbations which give origin to the CMB anisotropies. The relevance of these investigations has risen in light of the current and future experiments searching for GWs signature like  Ligo, Virgo and Kagra collaborations \cite{ligovirgo}, LISA \cite{Audley:2017drz}  Decigo \cite{dec1}, CE \cite{Evans:2016mbw}, Einstein Telescope \cite{ET-2}, just to name a few.

Apart from the interest of  having GWs which are intrinsically non-Gaussian, the non-linear nature of the source poses immediately one problem arising  from the fact that, while tensor modes are gauge invariant at first-order, they fail to remain so at second-order in perturbation theory. 
This point has been already noticed recently in the literature \cite{H,sasa,Gong:2019mui,Tomikawa:2019tvi}. 

In this paper we investigate the issue of the gauge invariance of the GWs. There are three steps to care of.  GWs have to be rendered gauge invariant at the production, during propagation and at the measurement. We will describe how to do so for each step.  There is not a single way to render the tensor modes gauge invariant at second-order.  Which gauge to use should be  in fact dictated by the measurement procedure, which we will describe.

 Let us elaborate on this point.
In cosmology one can build up  gauge invariant definitions of   physically defined, that is unambiguous, perturbations.
One should remember that there is a difference between  objects  which  are automatically gauge independent, i.e., they  have no gauge dependence (for instance a perturbation about a constant scalar), and objects  which are in general gauge dependent (think about the curvature perturbation) but can be rendered gauge invariant,  { in practice, by defining a combination which is truly gauge independent and coincide with that quantity in a particular gauge, see for example the discussion in Ref. \cite{Malik:2008im} (for instance the gauge invariant curvature perturbation $\zeta$ corresponding to  curvature perturbations on
 uniform density slices)}.
   Said alternatively,  gauge invariant quantities,  which do not depend on the   coordinate  definition of the perturbations in the given gauge, can be defined, and this is obtained in practice   by  unambiguously defining a given slicing into spatial hypersurfaces. 
For instance, the  tensor metric perturbation at linear-order  is gauge independent since it remains  the same
in all gauges, while on the contrary,  the gravitational potential is  gauge dependent since it  varies in different  time slicing.    
 
 A gauge invariant combination can be constructed, but it is not unique. 
This implies that there is  an infinite number of ways of making a quantity gauge invariant. Which is the best gauge one should start from  to compute the actual observables  is a matter that  can only be decided once the specifics of the measurement are understood. Once an observable is well-defined, there should not be any dependence on the gauge.

When dealing with the measurement of the GWs, in order to give a description of the response of the detector, the best choice  seems to be  the so-called TT frame \cite{book} which we will define and motivate in the following. 
Indeed the projected sensitivity curves for the interferometer LISA  are provided   in such a frame\footnote{We thank M. Maggiore for discussions about this point.}. From this point of view, in analogy with flat spacetime calculations, in the absence of a well-defined observable, the most reasonable gauge to choose is the TT gauge.
 Fortunately, once the GWs are produced and propagate  inside the horizon   to the detector, they can be treated as linear perturbations of the metric
 and,  as such,  they are gauge invariant. One expects therefore that the abundance of the GWs to be independent of the gauge. We are going to show it by comparing the result in the TT and in the Poisson gauges in the case in which the source is the one computed in the PBH scenario.

The paper is organised as follows. In Section 2 we discuss the measurement of the GWs, where we will follow Ref. \cite{book} and argue that the TT gauge turns out to be  the  preferred one from the practical point of view. In Section 3 we will discuss the gauge invariance of GWs at second-order in perturbation theory, and devote Section 4 to the gauge invariant expressions of the equations of motion, splitting the discussion into two parts, emission and propagation. Section 5 contains the computation of the abundance of GWs  while Section 6 contains our conclusions. Three  Appendices are devoted to summarise various technicalities the reader may find useful.

\section{The Measurement of GWs}\label{secm}
\noindent
In this section we discuss how GWs are observed and what the experimental apparatus is able to measure. 

We follow the steps described in Ref.~\cite{book} where many more details can be found. 
The measurement of GWs takes place in time intervals which are much smaller than the typical rate of change of the cosmological background, thus one can neglect the expansion of the universe and work with an (approximately) flat spacetime. In other words, we can take the flat spacetime limit, that is the limit in which we can put the scale factor $a=1$.

We focus our attention to experiments devoted to the measurement of GWs using interferometers. We simplify the discussion assuming the two arms A and B of lengths $\LA \sim \LB \sim L$ are aligned in the $\hat x$ and $\hat y$ directions, respectively. We also fix the origin of our coordinate system with the position of the beam splitter at the initial time $t_0$. 
In simple terms, the measurement is performed by sending a bunch of photons to the mirrors and measuring the modulation in power recorded back to the receiver due to the different time shifts $\Delta t_{\text{\tiny A},\text{\tiny B}}$ acquired in the different travel paths. For sake of simplicity, let us consider
a given component of the electric field vector (of frequency $\omega_\text{\tiny L}$) which gets a phase shift in both arms given by
\begin{align}
	 E_\text{\tiny A} (t) = -\frac{1}{2} E_0 {\rm e}^{-i \omega_\text{\tiny L} \lp t-2 \LA \rp + i \Delta \phi_\text{\tiny A}(t)}
	 \qquad \text{with}\qquad 
	 \Delta \phi_\text{\tiny A} = -\omega_\text{\tiny L} \Delta t_\text{\tiny A},
	 \\
	 E_\text{\tiny B} (t) = -\frac{1}{2} E_0 {\rm e}^{-i \omega_\text{\tiny L} \lp t-2 \LB  \rp + i \Delta \phi_\text{\tiny B}(t)}
	 \qquad \text{with}\qquad 
	 \Delta \phi_\text{\tiny B} = -\omega_\text{\tiny L} \Delta t_\text{\tiny B}.
\end{align}
What the measurement is actually able to observe is the total power of the electric field $P \sim |E_\text{\tiny A}+E_\text{\tiny B}|^2$ which is modulated by the GW as
\begin{equation}
P(t) = \frac{P_0}{2} \left \{ 1-\cos \llp 2 \phi_0 + \Delta \phi_\GW (t)\rrp\right \},
\end{equation}
where we have conveniently defined 
$\phi_0 = k_\text{\tiny L} (\LA - \LB)$ and $\Delta \phi_\GW (t)=\Delta \phi_\text{\tiny A} -\Delta \phi_\text{\tiny B}$.
In general, the passage of the GW can induce a time shift in two ways which are frame dependent. One is the movement of the mirrors (which is described by its geodesic motion) while the other is the change of the photon geodesic.

Dealing with interferometers, there are two kind of observatories which are currently used and planned to measure GWs, the space-based detectors and the ground-based ones. 
Thanks to the Equivalence Principle, in a small enough region it is always possible to choose the Fermi normal coordinates such that  the metric is flat. 
However, corrections to the flat metric arise starting quadratically in the ratio $(L /L_\text{\tiny BG})$ and $(L/\lambda_{\GW})$, where $L_\text{\tiny BG}$ identifies the typical length scale of variation of the background and $\lambda_{\GW}$ the GW wavelength, such that one gets
	\begin{equation}\label{properdetmetric}
		\d s^2 = -\d t^2 \lp 1+ R_{i0j0} x^i x^j \rp  -2  \d t \d x^i \lp \frac{2}{3} R_{0ijk} x^j x^k \rp + \d x^i \d x^j \llp  \delta_{ij} -\frac{1}{3} R_{ijkl} x^k x^l \rrp + \dots 
	\end{equation}
While for space-based detectors the mirrors are free falling and the corrections from Eq.~\eqref{properdetmetric} arise from the passage of the GW, for ground-based experiments one has to deal with the Earth gravity and the fact that one is fixed with a non-inertial frame. Therefore one gets additional contributions proportional to the local acceleration $a^i$ and angular velocity $\Omega^i$ of the laboratories with respect to the local gyroscopes. The effects are the inertial acceleration (2 ${\bf a}\cdot \bm{x}$), the gravitational redshift $({\bf a}\cdot \bm{x})^2$, the Lorentz time dilation due to the rotation of the laboratory $({\bf \Omega} \times \bm{x})^2$ and the so called ``Sagnac effect'' (${\bf \Omega} \times \bm{x}$). The study and characterisation of such effects is an experimental challenge and gives rise to the shape of the noise curves along with all the other relevant instrumental contributions. More broadly, the frame in which the metric takes the form \eqref{properdetmetric} is called the proper detector frame.

There are nonetheless more fundamental differences between these two apparatus which are due to the relation between the size of the arms and the characteristic frequency of the GWs. For ground-based detectors the typical GW frequency is of the order of $\omega_\GW L \sim 10^{-2}$, which allows to describe the effect of the GW in a Newtonian sense as a force acting on the mirrors  described by their geodesic deviation equation. 
For space-based observatories like LISA one has $\omega_\GW L \sim \pi/2$ and it is not possible to define a single reference frame where the whole apparatus is described by an (approximately) flat metric in the presence of the GW. Thus one is forced to work in a completely general relativistic framework where the most suitable coordinate system is the TT frame (known as the synchronous frame in cosmology). In simple words, the TT frame is defined by setting the coordinates in the positions of the mirrors. 

In the next sections we are going to review the GWs effect in the proper detector frame and in the TT frame.  Following the aforementioned arguments, notice that the latter is the most suitable one for the computation of the projected sensitivity curves for the LISA experiment.

\subsection{The measurement in the proper detector frame at first-order}
\noindent
If one assumes that $\omega_\GW L \ll 1$,  the system can be described in the proper detector frame. In this frame, the photons travel through a locally flat spacetime region while the mirrors are moved by the GWs.
Since an object at rest acquires a velocity $\d x^i / \d \tau \sim {\cal O} (\delta g)$, one has
\be
\d t^2 = \d \tau^2 \lp 1+ \frac{\d x^i}{\d \tau}\frac{\d x^i}{\d \tau} \rp =\d \tau^2 \lp 1+ \mathcal{O}(\delta g^2) \rp,
\ee
and time intervals correspond to proper time intervals.
This is a consequence of the fact that we are able to define a single reference frame in which the metric is (approximately) flat encompassing the whole apparatus (up to corrections of the order $\omega_\GW L \ll 1$).

Assuming that the detector is not relativistic, i.e. its velocity is small with respect to $c=1$ and $\d x^i / \d \tau \ll \d x^0 / \d \tau $, then one can write down the equation of motion for the infinitesimal  displacement of the mirrors as
\be
\frac{\d^2\xi^i}{\d \tau^2} = - R^{i}_{0j0} \xi^j  \lp \frac{\d x^0}{\d \tau} \rp^2.
\ee
For simplicity, we  restrict ourselves to the displacement equation at $\mathcal{O}(\delta g)$, with $R^i_{0j0}$ induced by the GW which is already at $\mathcal{O}(\delta g)$, and one can identify $t=\tau$ to get
\begin{equation}
\ddot \xi_i = - R_{i0j0} \xi^j,
\end{equation}
where the dot denotes the derivative with respect to the coordinate time $t = \tau$ of the proper detector frame. One can solve this equation perturbatively at first-order in $\delta g$ by noticing that
\begin{equation}
	\xi_\text{\tiny A} ^i = \llp \LA  + \delta \xi_\text{\tiny A}  +  {\cal O}(\delta g^2) \rrp \hat x^i
	\qquad \text{and} \qquad
	\xi_\text{\tiny B} ^i = \llp \LB  + \delta \xi_\text{\tiny B}  +  {\cal O}(\delta g^2) \rrp \hat y^i
\end{equation}
finding
\begin{align}
	\delta \xi_\text{\tiny A} (t)= - L_\text{\tiny A} \int^t \d t' \int^{t'} \d t'' R_{1010}
	\qquad \text{and} \qquad
	\delta \xi_\text{\tiny B} (t)= - L_\text{\tiny B} \int^t \d t' \int^{t'} \d t'' R_{2020}.
\end{align}
Following the  notation defined in appendix A for the generic metric, 
which we rewrite here for the convenience of the reader, 
\begin{align}
	\label{metric}
	\d s^2 =-a^2(1 +2  \phi) \d \eta^2+ 2 a^2 B_i \d \eta \,\d x^i +a^2 (\delta_{ij}+ 2 C_{ij}) \d x^i \d x^j,
\end{align}
 one can write in the flat spacetime limit
 the Riemann tensor $R_{i0j0}^{(1)}$ at first-order  
 in full generality as 
\begin{equation}
	R^{(1)}_{i0j0} =   
	   \partial _{i}\partial _{j}\phi_1
	   +
   \partial _{(i}B_{1j)} '
    - C_{1ij} '',
\end{equation}
which is gauge invariant under first-order coordinate transformations, see Eq.~\eqref{gt1o}.
It can be recasted in a more explicit form in terms of quantities which are individually gauge invariant in the flat spacetime limit as \cite{Flanagan:2005yc,book}
\begin{equation}
	R^{(1)}_{i0j0} =    -\frac{1}{2} h_{1ij}''  + \Delta_{,ij} + \partial_j \Xi'_i +  \partial_i \Xi'_j -\frac{1}{2} \Theta'' \delta_{ij},
\end{equation}
where
\begin{align}
\Delta =  \lp \phi_1 - E''_1 +  B' _1\rp   ,
\qquad
\Theta = -2 \psi_1  ,
\qquad
\Xi_i =  -\frac{1}{2}  \lp F_{1i}'+S_{1i} \rp .
\end{align}
 Such a property allows us to compute the Riemann tensor in any reference frame.
 In general, one can find the proper time it takes for a photon to complete a round trip ($\d t = \pm \d x$) to get
 \begin{equation}\label{pdfr}
 	\Delta t_{\text{\tiny A},\text{\tiny B}} = 2 \int_{t_0} ^{L_{\text{\tiny A},\text{\tiny B}} + \delta \xi_{\text{\tiny A},\text{\tiny B}}} \d x = 2 L_{\text{\tiny A},\text{\tiny B}} + 2 \delta \xi_{\text{\tiny A},\text{\tiny B}}(t_0+L_{\text{\tiny A},\text{\tiny B}}).
 \end{equation}
 The gauge invariance of the time shift at first-order is related to the fact that, in this frame, the coordinate time interval equals the proper time one.

We will show in the following that such a description turns out to be equivalent to the one in the TT frame at first-order in $\omega_\GW L \ll 1$.

\subsection{The measurement in the TT frame at second-order}
\begin{figure}[t!]
\centering
\includegraphics[width=0.45\columnwidth]{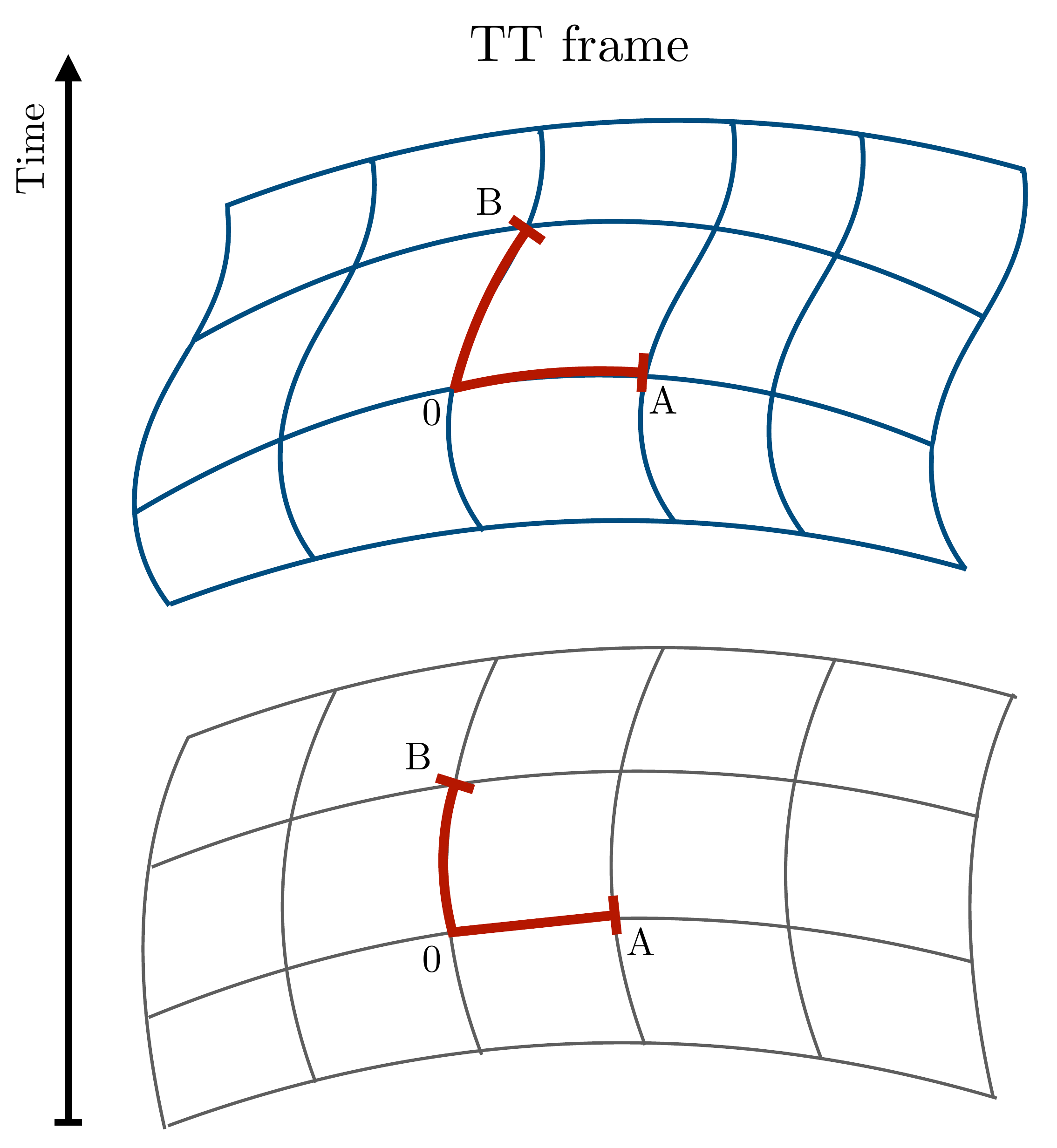}
\caption{Pictorial representation of the physical definition of the TT frame. The coordinates in such a frame are chosen such that the positions of the interferometer arms (in red) do not move even in the presence of a GW.
}
\label{schema1}
\end{figure}

\noindent
The movement of a test mass  in a general curved background is described by the geodesic equation of motion as
	\begin{equation}\label{geodesic}
\frac{\d^2 x^\mu}{\d \tau^2} = -\Gamma^\mu_{\nu\rho}  \frac{\d x^\nu}{\d \tau}\frac{\d x^\rho}{\d \tau},
\end{equation}
where $\tau$ identifies the proper time. Assuming the test mass to be at rest at the initial time $\tau_0$, the spatial components of the geodesic equation are
\begin{equation}
\frac{\d^2 x^i}{\d \tau^2} = -\Gamma^i_{00} \lp \frac{\d x^0}{\d \tau} \rp^2.
\end{equation}
This equation greatly simplifies by realising that the Christoffel symbols vanish, at any order in perturbation theory,
if one goes to the TT gauge, which is characterised by $\delta g_{00}=0$ and $\delta g_{0i}=0$. 
Following the  notation of  the generic metric in Eq. \eqref{metric}, 
the TT gauge corresponds to set
\begin{equation}\label{syncgauge}
\wt	\phi = 0
	\qquad \text{and} \qquad 
\wt	B_i = 0.
\end{equation}
As we already mentioned, this gauge is also referred to as the ``synchronous gauge'' in cosmology.
In such a gauge the test masses remain at rest with respect to the coordinates $x^i$. In other words, this corresponds to fixing the coordinates with the positions of the mirrors, see Fig.~\ref{schema1}.

The physical effect of a passing GW is captured by studying the proper times which are measured at the interferometers. Once emitted, the photons travel along the arms following the geodesic equation 
\begin{equation}
	\d s^2 = 0=
	- \d t^2 
	+ \lp \delta_{ij} +2 C_{1ij} + C_{2ij} \rp\d x^i \d x^j.
\end{equation}
Focusing, for example, on the arm A of the detector, at second-order one finds
\begin{equation}
	\d x = \pm \d t \llp 1 - {C_1}_{ij}  +\frac{3}{2}{C_1}_{ij}^2 - \frac{1}{2} {C_2}_{ij} \rrp_{i=j=1}+\dots
\end{equation}
where the upper sign holds for the travel towards the mirror and the lower one for the travel back to the beamsplitter. In general, the time shift up to second-order in the TT gauge takes the form 
\begin{tcolorbox}[colframe=white,arc=0pt]
		\vspace{-.3cm}
\begin{align}
	\label{tdTT}
	\Delta t_{\text{\tiny A},\text{\tiny B}} & = L_{\text{\tiny A},\text{\tiny B}} - \int_{t_0}^{t_0+2 L_{\text{\tiny A},\text{\tiny B}}}
	\d t
	 \lp - {C_1}_{ij}  +\frac{3}{2}{C_1}_{ij}^2 - \frac{1}{2} {C_2}_{ij} \rp _{i=j=1,2}.
\end{align}
\end{tcolorbox}
\noindent
Limiting ourself to the first-order in perturbations $\delta g$ and expanding at leading order in $\omega_\GW L\ll 1$ one finds
\begin{equation}
	\Delta t_{\text{\tiny A},\text{\tiny B}}  = L_{\text{\tiny A},\text{\tiny B}} 
	\lp 1+ 2 {C_1}_{ij}  \rp_{i=j=1,2} .
\end{equation}
It is not a surprise that, using the result in the proper detector frame and employing the gauge invariance of the Riemann tensor at first-order to evaluate it in the advantageous TT gauge, one recovers the same result for the time shift, see Eq.~\eqref{pdfr}. This is the manifestation of the fact that, assuming slowly varying perturbation fields, the time shift is a gauge invariant quantity as we will describe in details in the following.

\subsection{The measurement in a general frame at first-order}
\noindent
The GW affects the time shifts in two ways,  one is the change of the geodesic equation for the photon path; the other is the change of coordinate position of the mirrors. One can see that the computation in the TT gauge fixes the latter and all the physical effect is obtained via the photon geodesic equation. On the contrary in the proper detector frame, one fixes the photon geodesic, which is given by the propagation of the photon in (approximately) flat spacetime in the Fermi coordinates frame, and the GW impacts the position of the mirrors. 
In a general frame one needs to take into account both effects perturbatively, which we want to do in this subsection.

Let us start with the geodesic equation given by $\d s^2=0$ with
\begin{equation}
	\d s^2 = 
	-\lp 1+2 \phi\rp  \d t^2 
	+ 2B_{i}  \d x^i \d t
	+\lp \delta_{ij} +2 C_{ij}  \rp\d x^i \d x^j
\end{equation}
and, in the $x$-arm for example, one finds
\begin{equation}
	\d x= \pm \d t \lp 1 +\phi(t)\mp B_1(t) -C_{11}(t)\rp.
\end{equation}
The time it takes to the photon to arrive at the mirror (which is generally at position $L_{\text{\tiny A}} + \delta\xi_x$) is given by
\begin{equation}
t_1-t_0 = 	L_{\text{\tiny A}} +  \delta\xi_x(\delta g) -\int_{t_0}^{t_0+L_{\text{\tiny A}}}  \d t \llp \phi(t)- B_1(t) -C_{11}(t)\rrp.
\end{equation}
The remaining piece $\delta \xi_x(\delta g)$ is the one coming from the movement of the mirror and can be computed studying the geodesic  deviation equation in a general reference frame shown in Eq.~\eqref{geodesic}. We find 
\begin{equation}
	\delta 	\ddot \xi_x=
	- \lp B_1^{\prime} + \partial_1\phi \rp
\end{equation}
 at first-order in $\delta g$, whose solution is
\begin{equation}
	 \delta \xi_x =  -  \int^t_{t_0} \d t' \int_{t_0}^{t'} \d t'' \llp B_1^{\prime} (t'')+ \partial_1\phi (t'')\rrp,
\end{equation}
where the constants have been set requiring an oscillatory movement of the mirror and zero shift at $t_0$.

The total time shift of the photon in each arm is then given by
\begin{align}
	\label{gendt}
\Delta t_{\text{\tiny A},\text{\tiny B}} & =
2 L_{\text{\tiny A},\text{\tiny B}} 
+ \llp \int^{t_0+2L_{\text{\tiny A},\text{\tiny B}}}_{t_0} \d t' \int_{t_0}^{t'} \d t'' \llp B_i^{\prime} (t'')+ \partial_i\phi (t'')  \rrp \bigg |_{x_i=0}
-2 \int^{t_0+L_{\text{\tiny A},\text{\tiny B}}}_{t_0} \d t' \int_{t_0}^{t'} \d t'' \llp B_i^{\prime} (t'')+ \partial_i\phi (t'') \rrp \bigg |_{x_i=L_{\text{\tiny A},\text{\tiny B}}} \right.
\nn \\
& \left. -\int_{t_0}^{t_0+L_{\text{\tiny A},\text{\tiny B}}}  \d t \llp \phi(t)- B_i(t) -C_{ij}(t) \rrp
-\int_{t_0+L_{\text{\tiny A},\text{\tiny B}}}^{t_0+2 L_{\text{\tiny A},\text{\tiny B}}}  \d t \llp \phi(t)+ B_i(t) -C_{ij}(t) \rrp \rrp_{i = j = 1,2}.
\end{align}
Setting the TT gauge, one would recover the result found in Eq. \eqref{tdTT} at first-order in perturbation theory. 

In the limit in which the scalar and vector perturbations do not change considerably during the travel path of the photon, the time shift becomes
\begin{align}
\Delta t_{\text{\tiny A},\text{\tiny B}} & =
2 L_{\text{\tiny A},\text{\tiny B}}  -\int_{t_0}^{t_0+2 L_{\text{\tiny A},\text{\tiny B}}}  \d t \llp \phi(t) -C_{ij}(t)\rrp_{i = j = 1,2},
\end{align}
where we considered the leading order expansion in derivatives of the perturbation fields.
One can perform a gauge transformation such that the time shift transforms as
\begin{align}
\delta \llp \Delta t_{\text{\tiny A},\text{\tiny B}} \rrp &  = -\int_{t_0}^{t_0+2 L_{\text{\tiny A},\text{\tiny B}}}  \d t  \alpha'_1  = - \llp \alpha_1(t_0+2 L_{\text{\tiny A},\text{\tiny B}}) -	\alpha_1( t_0) \rrp,
\end{align}
where we have assumed that   $C_{ii}$ is gauge invariant at first-order in the limit of slowly varying perturbations in spacetime, thus finding the expected transformation property. It is then possible to define a gauge invariant quantity by considering
\vspace{-.0cm}
\begin{equation}
	\Delta t^\GI_{\text{\tiny A},\text{\tiny B}} \equiv \Delta t_{\text{\tiny A},\text{\tiny B}} + \int_{t_0}^{t_0+ 2 L_{\text{\tiny A},\text{\tiny B}}}   \phi (t) \d t,
\end{equation}
\noindent
which can be seen physically by realising that the change in time of the lapse function is slow compared to the time scale of the measurement, and therefore one can redefine a local time variable where the lapse function is absorbed in the new time coordinate seen by a local observer $\d \bar t \to (1+ \phi)  \d  t$. One recognises the so-called "gravity gradient noise" due  to the Newtonian gravitational potential evaluated at the extrema of the arms of the interferometer \cite{book}.

One can finally highlight the fact that such a definition of the gauge invariant time interval $\Delta t^\GI_{\text{\tiny A},\text{\tiny B}}$ corresponds to the solutions found in the TT frame at first-order
\begin{equation}
\Delta t^\GI_{\text{\tiny A},\text{\tiny B}} =	2 L_{\text{\tiny A},\text{\tiny B}}  +\int_{t_0}^{t_0+2 L_{\text{\tiny A},\text{\tiny B}}} \d t \,  C_{ij}(t)\Big|_{i = j = 1,2}.
\end{equation}

\section{Gauge invariant second-order tensor perturbations}
\noindent
There exists a precise and simple prescription on how to define gauge invariant quantities out of the quantities computed in a specific gauge \cite{Malik:1998ai,Malik:2003mv,ngreview,Malik:2008im,koyama}. As we argued in the introduction, building a gauge invariant combination only accounts for fixing the coordinate dependence of quantities  and provides the possibility of working with explicitly gauge invariant quantities, while it does not address the question of what is the physical observable, which is tightly related to the nature of the measurement performed.

The procedure is the following. One start by considering a certain gauge. In practice, one performs a coordinate transformation of the form
\begin{equation}
	x^\mu \to \wt x^\mu = x^\mu + \xi^\mu
	\qquad \text{with} \qquad 
	\xi^\mu \equiv \lp \alpha, \xi^i\rp.
\end{equation}
This fixes the gauge parameters $\xi^\mu$ one needs to use to reduce all the expressions to that particular gauge. In other words, the parameters $\alpha_{1}^\text{\tiny GC}$ and $\xi_{1i}^\text{\tiny GC}$ which enforce the gauge conditions can be expressed in terms of the perturbation fields (or combination thereof). Then, these particular combinations $\alpha_{1}^\text{\tiny GC} (\delta g)$ and $\xi_{1i}^\text{\tiny GC}(\delta g)$ can be used to perform a general gauge transformation to the original fields to find the gauge invariant quantities. Let us stress that the combination of fields obtained is explicitly gauge independent and defined regardless of the choice of any gauge.

Let us show this procedure for the case of the first-order  scalar potentials $\phi_1$ and $\psi_1$.
Performing a gauge transformation with parameters $\alpha_{1}^\text{\tiny GC} (\delta g)$ and $\xi_{1i}^\text{\tiny GC}(\delta g)$, then one obtains the gauge invariant scalar perturbations using Eq.~\eqref{gt1o} as
\begin{align}\label{sgi1}
 \phi_{1}^\GI & \equiv \phi_{1}  + \H \alpha_1^\text{\tiny GC}+\alpha_1^{\text{\tiny GC} \prime},
  \\
\label{sgi2}
 \psi_1^\GI &\equiv
\psi_1 - \H \alpha_1^\text{\tiny GC}.
\end{align}
One can check that such combinations are explicitly gauge invariant.

The same procedure can be used to define gauge invariant
second-order tensor perturbation. Using the gauge transformation properties of the tensor as in  Eq.~\eqref{transhij2} one defines \cite{Matarrese:1997ay,Malik:1998ai}\footnote{One may be surprised by the presence of non-local terms in the definition of the gauge invariant second-order tensor modes. They are present to ensure that the modes are transverse and traceless. However, these terms disappear in the "projected" equation of motion.}
\begin{align} 
h_{2ij}^\GI& \equiv h_{2ij}+\X_{ij}^\GC
+\frac{1}{2}\left(\nabla^{-2}\X^{\GC kl}_{~~~,kl}-\X^{\GC k}_{~~~k}
\right)\delta_{ij}
+\frac{1}{2}\nabla^{-2}\nabla^{-2}\X^{\GC kl}_{~~~~,klij}
+\frac{1}{2}\nabla^{-2}\X^{\GC k}_{~~~k,ij}
-\nabla^{-2}\left(\X_{ik,~j}^{\GC k}+\X_{jk,~i}^{\GC k}
\right),
\end{align}
where
\begin{align}
\X_{ij}^\GC &\equiv 
2\Big[\left(\H^2+\frac{a''}{a}\right)\alpha_1^{\GC 2}
+\H\left(\alpha_1^\GC\alpha_1^{\GC \prime }+\alpha_{1,k}^\GC \xi_{1}^{\GC k}
\right)\Big] \delta_{ij}\nonumber\\
&
+4\Big[\alpha_1^\GC \left(C_{1ij}'+2\H C_{1ij}\right)
+C_{1ij,k}\xi_{1}^{\GC k}+C_{1ik}\xi_{1~~,j}^{\GC k}
+C_{1kj}\xi_{1~~,i}^{\GC k}\Big]
+2\left(B_{1i}\alpha_{1,j}^\GC +B_{1j}\alpha_{1,i}^\GC \right)
\nonumber\\
&
+4\H\alpha_1^\GC \left( \xi_{1i,j}^\GC +\xi_{1j,i}^\GC \right)
-2\alpha_{1,i}^\GC \alpha_{1,j}^\GC +2\xi_{1k,i}^\GC \xi_{1~~,j}^{\GC  k}
+\alpha_1^\GC \left( \xi_{1i,j}^{\GC  \prime}+\xi_{1j,i}^{\GC  \prime} \right)
+\left(\xi_{1i,jk}^\GC +\xi_{1j,ik}^\GC \right)\xi_{1}^{\GC k}
\nonumber\\
&+\xi_{1i,k}^\GC \xi_{1~~,j}^{\GC  k}+\xi_{1j,k}^\GC \xi_{1~~,i}^{\GC  k}
+\xi_{1i}^{\GC \prime}\alpha_{1,j}^\GC +\xi_{1j}^{\GC \prime} \alpha_{1,i}^\GC 
\end{align}
in terms of the fields $\alpha_{1}^\text{\tiny GC} (\delta g)$ and $\xi_{1i}^\text{\tiny GC}(\delta g)$.
Notice that, in principle, one can construct different gauge invariant  quantities by using this procedure starting from different gauges.

When dealing with the equation of motion in momentum space, it is useful to introduce 
\begin{equation}
		 h_\lambda(t,\bm{k}) = e_\lambda^{ij}(\bm{k}) \int \d^3x e^{-i\bm{k}\cdot\bm{x}} h_{ij}(t,\bm{x}),
\end{equation}
where the polarisation tensor $e_\lambda^{ij}(\bm{k})$ is defined in appendix~\ref{appA}. Therefore, one can see that $ h_\lambda^\GI$ is constructed at second order as
\begin{equation}
	h_{2\lambda}^\GI = h_{2\lambda} + e_\lambda^{ij}(\bm{k})\X_{ij}^\GC.
\end{equation}
As we stressed in the introduction, the construction of the gauge invariant tensor modes is not unique and we will provide an example in the following.

\subsection{Explicit construction from the Poisson gauge}

We clarify the meaning of the construction procedure highlighted above by showing the explicit example starting from the Poisson gauge. We chose this particular gauge for convenience since it is the one commonly used to solve for the GWs induced at second-order by large scalar perturbations.

First of all, let us define the Poisson gauge by requiring that
\begin{equation}
\wt E^\text{\tiny P}=0,
\qquad
\wt B^\text{\tiny P}=0
\qquad \text{and} \qquad 
\wt {S}_i^\text{\tiny P}=0.
\end{equation}
To sum up, using the gauge transformation property in appendix~\ref{app2}, at first-order the gauge fixing is completely specified by setting
\begin{equation}
\alpha_{1}^\text{\tiny P} =B_1 -E_1',
\qquad 
\beta_1^\text{\tiny P} = - E_1,
\qquad
\gamma_{1i}^{\text{\tiny P}}=\int^\eta  S_{1i} \d\eta' + \hat\C_{1i}(\bm{x}),
\end{equation}
up to an arbitrary constant 3-vector $\hat\C_{1i}$ which depends on
the choice of spatial coordinates on an initial hypersurface.

 Using the choices above together with Eqs.~\eqref{sgi1} and \eqref{sgi2}, the gauge invariant first-order scalar perturbations are defined as 
\begin{align}
\label{defphi1l}
\Phi_1 & \equiv \phi_{1}  + \H \alpha_1^\text{\tiny P}+\alpha_1^{\text{\tiny P} \prime} = \A1 + \H(B_1-E_1') + (B_1-E_1')', \\
\label{defpsi1l} 
\Psi_1 &\equiv
\psi_1 - \H \alpha_1^\text{\tiny P}=  \psi_1 - \H \left( B_1-E_1'
\right).
\end{align}
One can easily check that these combinations are explicitly gauge invariant and equivalent to the Bardeen potentials \cite{Bardeen:1980kt}. Also, these gauge invariant combinations reduce to the known results if one chooses the Poisson gauge. 

The same procedure can be used to define gauge invariant
second-order tensor perturbations. Using the gauge transformation properties of the tensor as in  Eq.~\eqref{transhij2} one defines
\begin{align} 
h_{2ij}^{\GI,\PP}& \equiv h_{2ij}+\X_{ij}^\PP
+\frac{1}{2}\left(\nabla^{-2}\X^{\PP kl}_{~~~,kl}-\X^{\PP k}_{~~k}
\right)\delta_{ij}
+\frac{1}{2}\nabla^{-2}\nabla^{-2}\X^{\PP kl}_{~~~,klij}
+\frac{1}{2}\nabla^{-2}\X^{\PP k}_{~~k,ij}
-\nabla^{-2}\left(\X_{ik,~~j}^{\PP ~k}+\X_{jk,~~i}^{\PP ~k}
\right)
\end{align}
in terms of
\begin{align}
\X_{ij}^\PP &\equiv 
2\Big[\left(\H^2+\frac{a''}{a}\right)\alpha_1^{\PP 2}
+\H\left(\alpha_1^\PP\alpha_1^{\PP \prime }+\alpha_{1,k}^\PP \xi_{1}^{\PP k}
\right)\Big] \delta_{ij}\nonumber\\
&
+4\Big[\alpha_1^\PP \left(C_{1ij}'+2\H C_{1ij}\right)
+C_{1ij,k}\xi_{1}^{\PP k}+C_{1ik}\xi_{1~~,j}^{\PP k}
+C_{1kj}\xi_{1~~,i}^{\PP k}\Big]
+2\left(B_{1i}\alpha_{1,j}^\PP +B_{1j}\alpha_{1,i}^\PP \right)
\nonumber\\
&
+4\H\alpha_1^\PP \left( \xi_{1i,j}^\PP +\xi_{1j,i}^\PP \right)
-2\alpha_{1,i}^\PP \alpha_{1,j}^\PP +2\xi_{1k,i}^\PP \xi_{1~~,j}^{\PP  k}
+\alpha_1^\PP \left( \xi_{1i,j}^{\PP  \prime}+\xi_{1j,i}^{\PP  \prime} \right)
+\left(\xi_{1i,jk}^\PP +\xi_{1j,ik}^\PP \right)\xi_{1}^{\PP k}
\nonumber\\
&+\xi_{1i,k}^\PP \xi_{1~~,j}^{\PP  k}+\xi_{1j,k}^\PP \xi_{1~~,i}^{\PP  k}
+\xi_{1i}^{\PP \prime}\alpha_{1,j}^\PP +\xi_{1j}^{\PP \prime} \alpha_{1,i}^\PP.
\end{align}
The explicit expression of $\X_{ij}^\PP$ can be found in the appendix in Eq.~\eqref{XPP}.

\subsection{Issues in the TT gauge}

As we discussed in Section 2, the TT gauge, also dubbed the synchronous gauge in the cosmological setting, is the one to be preferred when dealing with the concept of the measurement of the GWs. 


The reader should be aware that in the TT gauge, as it will be clear from the equations in the following, it is not possible to construct truly gauge invariant quantities because the time-slicing is not unambiguously defined and there exists a residual gauge freedom.  
Let us start from the gauge transformation which allows to go to  the  TT gauge. Starting from the definition in Eq.~\eqref{syncgauge}
and using Eqs.~(\ref{gt1o}), one finds
\begin{align}
\alpha_{1}^\syn
&=-\frac{1}{a}\llp \int a\phi_1 \d\eta-\C_1(\bm{x})\rrp,\label{sync1}\\
\beta_{1}^{\syn}&=\int\left( \alpha_{1}^{\syn}-B_1\right) \d\eta
+\hat\C_1(\bm{x}),\label{sync2} \\
 \gamma_{1i}^{\syn} &= \int S_{1i} \d\eta + \hat\C_{1i}(\bm{x}).
 \label{syn3}
 \end{align}
The determination of the time-slicing is fully  done once one 
fixes the two arbitrary scalar functions of the spatial coordinates $\C_1(\bm{x})$ and $\hat\C_1(\bm{x})$. Also, one has a constant 3-vector $\hat\C_{1i}$ which depends on
the choice of spatial coordinates on an initial hypersurface. 
The presence of such constants makes it impossible to define truly gauge invariant quantities from the conditions \eqref{syn3} \cite{Malik:2008im,martin}.

Of course, the residual gauge freedom in the TT gauge does not appear when considering real observables (see, for example, the discussion about this point in \cite{ber}). At first-order the measurement process shows this property explicitly. At second-order, for instance if  one wishes to measure the non-Gaussian nature of the GWs,  one would have to build up  appropriate observables for which the residual gauge modes should similarly disappear.

\section{Gauge invariant equation of motion for GWs}
\noindent
 The equation of motion for the transverse and traceless metric perturbation $h_{ij}$ at second-order  can be written as (see, for example, Ref.~\cite{Gong:2019mui})
\begin{equation}
\label{EoM}
 h''_\lambda(\eta,\bm{k}) + 2 \H h'_\lambda(\eta,\bm{k}) + k^2 h_\lambda(\eta,\bm{k}) 
=  2 a^2(\eta)  e^{ij}_\lambda(\bm{k}) \, s_{ij}(\bm{k}),
\end{equation}
where the polarisation tensor is defined in Eq.~\eqref{pol}.
The source at second-order appearing in the equation of motion is composed by three different structures, namely the scalar-scalar, scalar-tensor and tensor-tensor as 
\be
s_{ij} = s_{ij}^{(ss)}+s_{ij}^{(st)}+s_{ij}^{(tt)}.
\ee
The last piece can be safely neglected, being at higher order in the tensor modes.
The scalar-scalar term can be regarded as responsible for the emission, while the scalar-tensor as the dominant one regarding the propagation. 

In order to conveniently simplify the notation, we introduce the shear potential $\sigma_1 = E_1'-B_1$.
The explicit expression, without specifying any gauge, for the sources is 
 \cite{Gong:2019mui}
\begin{align}
\label{eq:nij-ss}
s_{ij}^{(ss)} 
= &
-\frac{1}{a^4} \frac{\d}{\d \eta} 
\Big[
a^2 \Big( 2\psi_1 \sigma_{1,ij} 
+ \psi_{1,i}\sigma_{1,j} 
+ \psi _{1,j}\sigma_{1,i} \Big)
\Big]
+ \frac{1}{a^2}
\lp 
3 \H \phi_1
+3\psi_1'
-\sigma_{1,kk}
\rp
\sigma_{1,ij}
-
\frac{1}{a^2}
 \Big(  4\psi_1 \psi_{1,ij}  +   3\psi_{1,i}
 \psi_{1,j}
\Big)
\nonumber\\
& 
+ \frac{1}{a^2}\sigma_1^{,k}{}_{,i}\sigma_{1,jk}
+ \frac{1}{a^2}  \Big[ 
2\phi_1   \sigma_{1,ij}' 
+ \H\phi_1
 \sigma_{1,ij} 
  + \phi_1 ' \sigma_{1,ij} 
- 2 (\phi_1-\psi_1)\phi_{1,ij} 
-  \phi_{1,i}\phi_{1,j} 
+ 2  \psi_{1,(i}\phi_{,j)}
\Big]\nonumber\\
&
+ 8\pi G  (\rho+P) v_{,i}v_{,j}, 
\end{align}
where $v$ is the scalar velocity potential, $G$ is the Newton's gravitational  constant and\footnote{
One can notice that the scalar-tensor source is not manifestly symmetric in the indices $\{i,j\}$ unless one takes advantage of the equation of motion for the first-order perturbation in Eq.~\eqref{eq:aniso}.
} 
\begin{align}
s_{ij}^{(st)} 
= & 
\frac{1}{2 a}\frac{\d}{\d \eta} \bigg[ 
 \frac{1}{a} {h}'_{1ij}   \phi_1
- \frac{2}{a} \bigg(  \psi_1 {h}'_{1ij} +\psi_1' h_{1ij} -  h_{1i}^k\sigma_{1,jk} \bigg)
+ \frac{\sigma_1^{,k}}{a} \Big(  h_{1ik,j}  + h_{1jk,i}  - h_{1ij,k} \Big)
\bigg]
\nonumber\\
& + \frac{3}{2} \frac{\H}{a^2} \bigg[
 {h}'_{1ij}   \phi_1
- 2 \bigg(  \psi_1 {h}'_{1ij} +\psi_1' h_{1ij} -  h_{1i}^k\sigma_{1,jk} \bigg)
+ \sigma_1^{,k}\Big(  h_{1ik,j}  + h_{1jk,i}  - h_{1ij,k} \Big)\bigg]
\nonumber\\
& + \phi_1 \frac{1}{2 a} \frac{\d}{\d\eta } \bigg(  \frac{1}{a} {h}'_{1ij}  \bigg)
- \frac{1}{2 a^2}\sigma_1^{,k}   {h}'_{1ij,k}  
+ \frac{1}{2 a^2}  {h}'_{1ij}  \lp 
3 \H  \phi_1
+3  \psi_1'
 - \sigma_{1,kk}
\rp
+ \frac{1}{2 a^2}\sigma_{1,i}{}^{,k}   {h}'_{1jk}  
- \frac{1}{2 a^2} \sigma_{1,j}{}^{,k}  {h}'_{1ik}  
\nonumber\\
& 
- \frac{1}{2 a^2} \bigg[
 2   h_{1i}^k  \phi_{1,jk}
+ \Big(  h_{1ik,j}  + h_{1jk,i}  - h_{1ij,k} \Big) \phi_1^{,k}
\bigg]
+ \frac{1}{2 a^2} \bigg[
 2 \Big( 2\psi_1 {h}_{1ij,kk} - h_{1j}^k\psi_{1,ik}
  + h_{1ij}  \psi_{1,kk} \Big)
  \nn \\
&- \frac{1}{2}\psi_1^{,k} \Big( h_{1ik,j} + h_{1jk,i} - 3h_{1ij,k} \Big)
    \bigg].
\label{eq:nij-st}
\end{align}
The equation of motion Eq. \eqref{EoM}, expanded at second-order by keeping all the perturbations of the metric, is obviously gauge invariant by construction. This can be also checked explicitly by employing the second-order gauge transformation reviewed in appendix~\ref{app2}.

Typically in the literature such equations have been analysed and solved in the Poisson (Newtonian) gauge.
For a radiation-dominated universe, where the pressure density $P =\rho /3$, the scalar-scalar source in this gauge is given by\footnote{From now on we use the first-order dynamical equations of motion for the scalar perturbations in the Poisson gauge which imply,  in the absence of anisotropic stress, $\phi_1= \psi_1$, see appendix~\ref{app3}.}
\begin{align}
s_{ij, \text{\tiny P}}^{(ss)} 
&= -\frac{1}{a^2(\eta)} 
\llp 4 \psi _1 \psi_{1,ij}+2 \psi_{1,i}\psi_{1,j} - \partial_i \lp \frac{\psi_1'}{\H} + \psi_1 \rp \partial_j \lp \frac{\psi_1'}{\H} + \psi_1 \rp \rrp,
\end{align}
which reproduces the scalar-scalar emission source used in the literature. 
Similarly,  the scalar-tensor source in the same gauge in a general FRWL  background is 
\begin{align}
\label{sstP}
s_{ij, \text{\tiny P}}^{(st)} 
&= \frac{1}{a^2(\eta)}  \llp 2 \psi_1 h_{1ij,kk}
-h_{1ij} \lp 2 \H \psi_1 ' + \psi_1 '' -\psi_{1,kk}\rp 
-2 \psi_{1,k} \lp h_{1k(i,j)} - h_{1ij,k}\rp 
- 2  \psi_{1,k(i}h_{1j)k} \rrp,
\end{align}
where the  first term exactly reproduces the source for the Shapiro time delay in the scalar-tensor component (see, for example,  \cite{Bartolo:2018rku} and references therein).

\subsection{Gauge invariant emission equation from the Poisson gauge}
In order to describe in a gauge invariant way the emission of the GWs at second-order, one can start from the equation of motion of the GWs with the scalar-scalar source Eq.~\eqref{eq:nij-ss} and identify the various gauge invariant combinations. This is a straightforward, but tedious,  procedure which can be performed starting from any gauge. In particular, starting from the Poisson gauge one obtains 
\begin{align}
&h^{\GI,\PP\prime\prime}_{2ij} +  2\H  h^{\GI,\PP\prime}_{2ij} -   h^{\GI,\PP}_{2ij,kk} 
=  
- 4 \mathcal{T}_{ij}^{lm} 
\llp 4 \Psi _1 \Psi_{1,lm}+2 \Psi_{1,l}\Psi_{1,m} - \partial_l \lp \frac{\Psi_1'}{\H} + \Psi_1 \rp \partial_m \lp \frac{\Psi_1'}{\H} + \Psi_1 \rp \rrp,
\end{align}
where we introduced the transverse and traceless projector $\mathcal{T}_{ij}^{lm} $, see Eq. \eqref{Tproj}.
In practice, this is the equation of motion solved in the literature when dealing with GWs produced by second-order scalar perturbations, and one can immediately realise that both  sides of the equation are individually gauge invariant.

The fact that the equations of motion can be written in a completely gauge invariant way does not solve the issues mentioned in the literature. Namely, other gauge invariant definitions of tensor modes, yield a different form of the gauge invariant equations of motion and, therefore, different naive predictions for the induced gravitational waves. In the end, one needs to identify the observable quantity. Then, one may find the gauge invariant variable that best describes it.

\subsection{Gauge invariant propagation equation from the Poisson gauge}
For the propagation, one can similarly start from the equation of motion for the second-order GWs with the scalar-tensor source in Eq.~\eqref{eq:nij-st} and identify the various gauge invariant combinations. Starting from the Poisson gauge one gets, extracting the leading term responsible for the Shapiro time delay
\begin{align}
	&h^{\GI,\PP\prime\prime}_{2ij} +  2\H  h^{\GI,\PP\prime}_{2ij} - h^{\GI,\PP}_{2ij,kk}  =  4  \mathcal{T}_{ij}^{lm}   \lp 2  \Psi_1 h_{1lm,kk} \rp,
\end{align}
where we neglected in the full scalar-tensor source of  Eq. \eqref{sstP} terms with lower derivatives in the tensor modes, being subdominant with respect to the Shapiro time delay term in the geometrical optics approximation.

\section{The GW power spectrum in the TT gauge}
\noindent
As we argued above, the TT gauge should be preferred when dealing with the issue of the measurement, at least at the linear order. As already stressed, this is also motivated by the fact that, for instance, the sensitivity curve for LISA is provided in the TT frame. 

The impossibility to construct gauge invariant quantities which reproduce the ones in the TT gauge, as shown above, seems to suggest that abandoning the gauge invariant formalism is necessary and one should compute the physical observable in the specific gauge.
However, as we shall see, the gauge dependence is lost in the late time observables since the GWs effectively become linear perturbations of the metric and, as such, gauge invariant.

\subsection{Linear solutions in the  TT  gauge}

Here,  for convenience, we provide the explicit relation between the degrees of freedom in the Poisson gauge and the TT gauge. We start from the Poisson gauge where the solution at linear level is widely used in literature.
Using the gauge transformation definitions in  Eq.~\eqref{gt1o}, one finds
\begin{align}
\phi_1^\TT &= 0,
\qquad
 \psi_1^\TT  = \psi_1^\text{\tiny P}-\H\alpha_1^\TT ,
 \qquad 
 B_1^\TT =0,
 \qquad
 E_1 ^\TT=\beta_1^\TT , 
 \nn \\
 S_{1i} ^\TT&= 0 , 
 \quad \
 F_{1i}^\TT = F_{1i}^\text{\tiny P} ,
\nn \\
 h_{1ij}^\TT &= h_{1ij}^\text{\tiny P}.
\end{align}
The general expression for the gauge parameters is given in Eqs. (\ref{sync1}) - (\ref{syn3}), and we get
the scalar function $\psi_1^\TT$ in Fourier transform as
\begin{equation}
\psi_1^\TT ({\bm k},\eta) = \psi_1^\PP ({\bm k},\eta) 
-  \H  \llp \frac{ 1}{a(\eta)}{\cal C}_1 ({\bm k})
-
\frac{1}{a(\eta)} \int^\eta a(\eta') \psi_1^\PP ({\bm k},\eta')  \d \eta'
\rrp,
\end{equation}
while the shear potential $\sigma_1$ appearing in the equation of motion becomes
\begin{equation}
\sigma_1^\TT({\bm k},\eta) = E_1^{\TT \prime} ({\bm k},\eta)= -\frac{1}{\H} \llp  \psi_1^\TT ({\bm k},\eta) - \psi_1^\PP ({\bm k},\eta) \rrp.
\end{equation}
Specialising the result to a radiation-dominated epoch, where $a \sim \eta$ and $\H = 1/\eta$, one finds\footnote{We have arbitrarily reabsorbed the contribution from the lower limit of the integral in the overall constant, named ${\cal C}$.}
\begin{align}
\psi_1^\TT ({\bm k},\eta) &= 
\frac{2}{3} \zeta( {\bm k} )3\llp \frac{j_1(z)}{z} - \frac{j_0(z)}{z^2}\rrp
-  \frac{{\cal C} ({\bm k}) }{\eta^2},
\\
\sigma_1^\TT ({\bm k},\eta) &= 
\frac{2}{3} \zeta( {\bm k} )3 \eta  \frac{j_0(z)}{z^2}
+  \frac{{\cal C} ({\bm k}) }{\eta},
\end{align}
where $\zeta( {\bm k} )$ is the comoving curvature perturbation and $z= k \eta/\sqrt{3}$. For details about the linear transfer function of  the scalar perturbation in the Poisson gauge see appendix \ref{app3}. The choice of the constants can be made by requiring a finite value of the perturbations in the super-horizon limit  $k \eta \to 0$ in accordance with \cite{Ma}, which sets ${\cal \C} (\bm{k}) = -6 \zeta (\bm{k}) / k^2$ to get
\begin{align}
\label{Tfpsi}
\psi_1^\TT ({\bm k},\eta) &= 
\frac{2}{3} \zeta( {\bm k} ) 3\llp \frac{j_1(z)}{z} - \frac{j_0(z)}{z^2}
+ \frac{1 }{z^2} 
\rrp \equiv  \frac{2}{3} \zeta( {\bm k} ) T_\psi(\eta,k),
\\
\label{Tfsigma}
\sigma_1^\TT ({\bm k},\eta) &= 
\frac{2}{3} \zeta( {\bm k} ) 3\eta  \llp \frac{j_0(z)}{z^2} -\frac{1}{z^2} \rrp  
\equiv \frac{2}{3} \zeta( {\bm k} ) \frac{\sqrt{3}}{k} T_\sigma(\eta, k).
\end{align}
Specialising the result to a matter-dominated epoch instead, where $\psi_1^\PP (\bm{k}, \eta) = 3 \zeta (\bm{k})/5$, $a \sim \eta^2$ and $\H = 2/\eta$, one finds
\begin{align}
\psi_1^\TT ({\bm k},\eta) &= 
 \zeta( {\bm k}),
\\
\sigma_1^\TT ({\bm k},\eta) &= 
-\frac{1}{5} \zeta( {\bm k} ) \eta.
\end{align}
One can explicitly check that in both the matter- and radiation-dominated epochs, the solutions found satisfy the equation of motion \eqref{eq:aniso} specialised to the TT gauge and in the absence of anisotropic stress.

\subsection{GWs emission in the TT gauge}

In this subsection we compute the GW abundance in the TT gauge.
The emission source in a general FRWL background in Eq.~\eqref{eq:nij-ss} takes the form
\begin{align}
s_{ij, \text{\tiny TT}}^{(ss)} 
=-\frac{1}{a^2(\eta)} \Big[
&
 \psi_{1} \psi_{1,ij} 
+   2  \psi_{1,(i}' \sigma_{1,j)}
-   \psi_{1} ' \sigma_{1,ij}  
+ \sigma_{1,ij}  \sigma_{1,kk}  
- \sigma_{1,ik} 
 \sigma_{1,jk} 
+ \frac{2}{\H '-\H^2}  \psi_{1,i} ' \psi_{1,j} ' 
\Big],
\end{align}
and therefore the equation of motion for the emission of the tensor fields in a radiation-dominated universe is given, in momentum space, by
\begin{tcolorbox}[colframe=white,arc=0pt]
	\vspace{-.3cm}
\begin{align}
h^{\TT\prime\prime}_{\lambda} (\eta, \bm k) +  2\H  h^{\TT\prime}_{\lambda} (\eta, \bm k) +k^2  h^{\TT}_{\lambda} (\eta, \bm k)
=
- 4 & e_{\lambda}^{ij}(\bm k) \Big[
 \psi_{1} ^{\TT} \psi_{1,ij} ^{\TT}
+   2  \psi_{1,(i} ^{\TT \prime} \sigma_{1,j)}^{\TT }
-   \psi_{1} ^{\TT \prime} \sigma_{1,ij} ^{\TT} 
\nn \\
&+ \sigma_{1,ij} ^{\TT} \sigma_{1,mm}  ^{\TT}
- \sigma_{1,im} ^{\TT}
\sigma_{1,jm} ^{\TT}
- \frac{1}{\H^2}  \psi_{1,i} ^{\TT \prime} \psi_{1,j} ^{\TT \prime} 
\Big] \equiv \mathcal{S}^{\TT}_{\lambda} (\eta, \bm k).
\end{align}
\end{tcolorbox}
\noindent
The method of the Green function yields the solution (following the notation in Ref. \cite{sm})
\be
\label{solh}
h^{\TT}_{\lambda} (\eta, \bm k) = \frac{1}{a(\eta)} \int^\eta \d \eta' G_{\bm k}(\eta, \eta') a(\eta') \mathcal{S}^{\TT}_{\lambda} (\eta', \bm k),
\ee
where the Green function $G_{\bm k}(\eta, \eta')$  in a radiation-dominated universe is given by
\be
G_{\bm k}(\eta, \eta') = \frac{{\rm sin}(k (\eta - \eta'))}{k} \theta (\eta- \eta'),
\ee
with $\theta$ the Heaviside step function, and the source in momentum space is 
\begin{align}
\mathcal{S}^{\TT}_{\lambda} (\eta, \bm k) &= -4 \int \frac{\d^3 p}{(2\pi)^3} e^{ij}_\lambda (\bm k)
\llp - p_i p_j \psi_{1} ^{\TT} (\bm p) \psi_{1} ^{\TT} (\bm k - \bm  p) 
+2 p_i  p_j \psi_{1} ^{\TT \prime} (\bm p) \sigma_{1} ^{\TT } (\bm k - \bm  p) 
+ p_i p_j \psi_{1} ^{\TT \prime}  (\bm k - \bm p) \sigma_{1} ^{\TT} (\bm  p)  \right. \nonumber \\
&  \left.
+  p_i p_j (\bm k - \bm p)^2 \sigma_{1} ^{\TT} (\bm p) \sigma_{1} ^{\TT} (\bm k - \bm  p) 
+  (p_i p_j (\bm p \cdot \bm k) - p_i p_j p^2) \sigma_{1} ^{\TT} (\bm p) \sigma_{1} ^{\TT} (\bm k - \bm  p) 
- \frac{1}{\H^2} p_i  p_j \psi_{1} ^{\TT \prime}  (\bm p) \psi_{1} ^{\TT \prime} (\bm k - \bm  p) 
\rrp,
\end{align}
where we used the transversality property of the polarisation tensors $k_j e^{ij}(\bm k) = 0$.
Defining the object
\be
e_\lambda (\bm k, \bm p) \equiv e_\lambda^{ij} (\bm k) p_i p_j,
\ee
and using the definitions of the transfer function of the perturbation fields $\psi$ and $\sigma$ in the TT gauge given by Eqs. \eqref{Tfpsi} and \eqref{Tfsigma}, the source becomes
\begin{align}
\mathcal{S}^{\TT}_{\lambda} (\eta, \bm k) = 
\frac{4}{9} \int \frac{\d^3 p}{(2\pi)^3} e_\lambda (\bm k, \bm p)  \zeta (\bm k) \zeta (\bm k - \bm p) f^{\TT}(p, |\bm k - \bm p|, \eta),
\end{align}
in terms of the function
\begin{align}
f^{\TT}(p, |\bm k - \bm p|, \eta) & \equiv  
-4 \bigg [ 
- T_\psi (\eta, p) T_\psi (\eta, |\bm k - \bm p|)
+2 \frac{\sqrt{3}}{|\bm k - \bm p|}  T_\psi' (\eta, p) T_\sigma (\eta, |\bm k - \bm p|)
+  \frac{\sqrt{3}}{p}  T'_\psi (\eta, |\bm k - \bm p|) T_\sigma (\eta, p)  
\nonumber \\
&
+  3 \frac{(\bm k- \bm p)^2}{p |\bm k - \bm p|}T_\sigma (\eta, p)  T_\sigma (\eta, |\bm k - \bm p|) 
+ 3 \frac{\llp (\bm p \cdot \bm k) -  p^2 \rrp }{p |\bm k - \bm p|} T_\sigma (\eta, p)  T_\sigma (\eta, |\bm k - \bm p|) 
- \frac{1}{\H^2} T'_\psi (\eta, p) T'_\psi (\eta, |\bm k - \bm p|)
\bigg ].
\end{align}
Collecting this expression in the solution of the equation of motion in Eq. \eqref{solh},
and using the fact that the emission takes place for few Hubble times after horizon crossing, well within the radiation-dominated phase\cite{sm}, 
the solution becomes
\begin{align}
h^{\TT}_{\lambda} (\eta, \bm k) &= 
\frac{4}{9} \int \frac{\d^3 p}{(2\pi)^3} \frac{1}{k^3 \eta} e_\lambda (\bm k, \bm p)  \zeta (\bm k) \zeta (\bm k - \bm p)
\llp \mathcal{I}^{\TT}_c(x,y) \cos (k \eta) +  \mathcal{I}^{\TT}_s(x,y) \sin (k \eta)  \rrp,
\end{align}
in terms of the  oscillating functions 
\begin{align}
\label{osci}
\mathcal{I}^{\TT}_c(x,y) & = \int_0^\infty \d u \, u (-\sin u) f^{\TT}(x,y,u), \nonumber \\
\mathcal{I}^{\TT}_s(x,y) & = \int_0^\infty \d u \, u (\cos u) f^{\TT}(x,y,u).
\end{align}
The latter can be  computed from
\begin{align}
\label{fTT}
f^{\TT}(x,y,u)  \equiv  
-4 & \bigg [ 
- T_\psi (u x) T_\psi (u y)
+2 \frac{\sqrt{3}}{k y}  T_\psi ' (u x) T_\sigma (u y)
+  \frac{\sqrt{3}}{k x}   T_\sigma (ux)  T'_\psi (u y)
\nonumber \\
& 
+ \frac{3}{2} \frac{(1- x^2 + y^2)}{x y} T_\sigma (ux )  T_\sigma (u y) 
- \frac{u^2}{k^2} T'_\psi (u x) T'_\psi (uy)
\bigg ],
\end{align}
where we introduced the dimensionless parameters $x = p/k$, $y = |\bm k - \bm p|/k$ and $u = k \eta'$. Its explicit expression can be found in Appendix \ref{appc}.

\begin{figure}[t!]
\centering
\includegraphics[width=0.49\columnwidth]{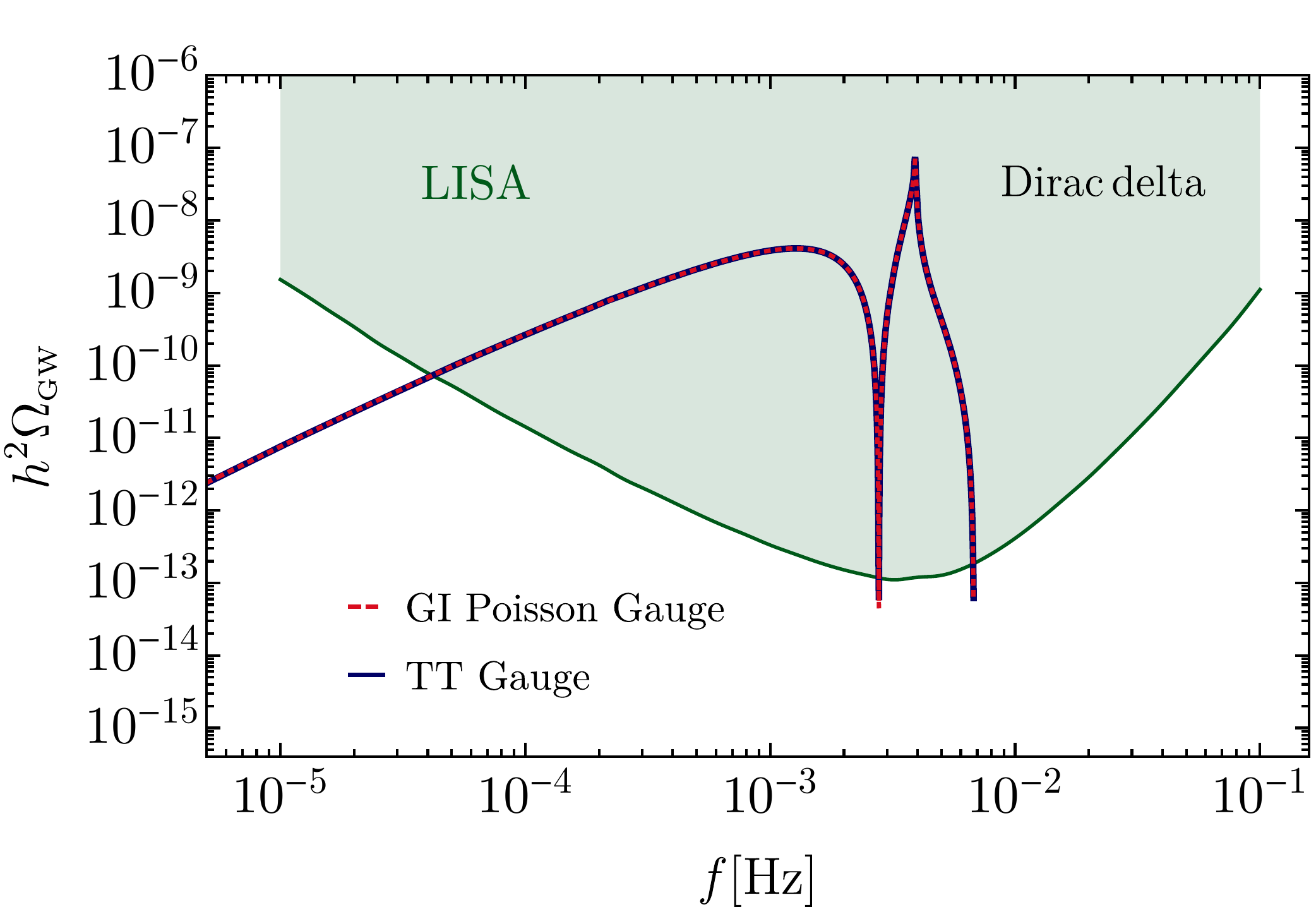}
\includegraphics[width=0.49\columnwidth]{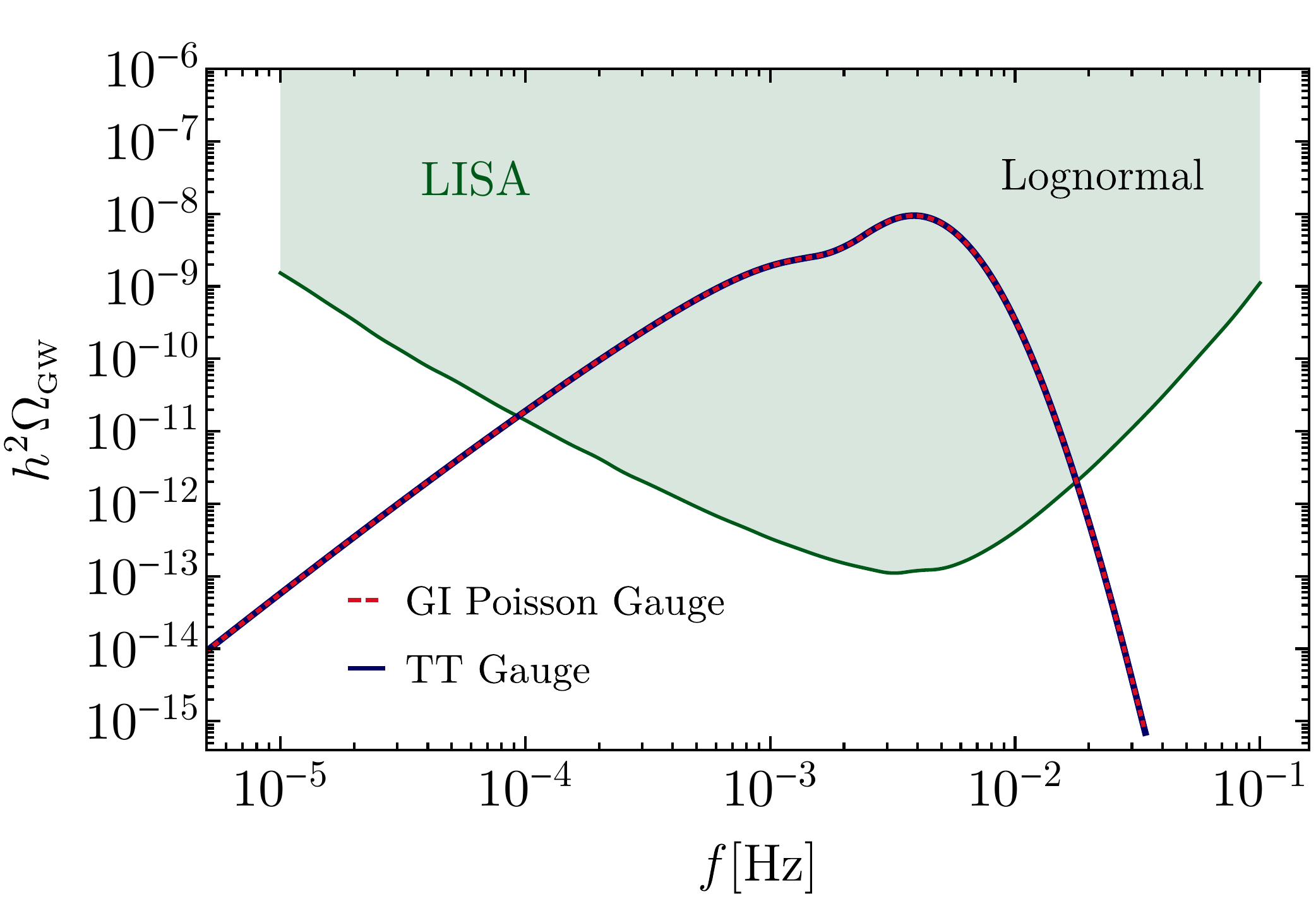}
\caption{
Plot of the abundances of GWs computed in the TT gauge (blue solid) and the one using the gauge invariant definition from the Poisson gauge (red dashed) along with the 
estimated sensitivity for LISA \cite{Audley:2017drz}. 
We have used the value $A= 0.033$ for the amplitude of the Dirac delta power spectrum (left panel), $A_s =0.055$ and  $\sigma = 1/2$ for the lognormal one (right panel) \cite{Bartolo:2018rku}. The characteristic   wavenumber  $k_* \equiv  2 \pi f_* =   21\, {\rm mHz}$ was chosen in order to have PBH with masses $10^{-12}M_\odot$ as the totality of dark matter.
}
\label{fig: Omega GW}
\end{figure}

The GW energy density is defined to be \cite{Maggiore:1999vm} 
\begin{equation}
\rho_\GW = \frac{1}{32 \pi G a^2} \left\langle {h}_{ij}' \left( \eta ,\, \bm{x} \right)  {h}_{ij} '\left( \eta ,\, \bm{x} \right) \right\rangle_\text{\tiny T},
\label{rhoGW}
\end{equation} 
where the label T denotes a time average over multiple various characteristic periods of the wave.
For a general comoving curvature perturbation power spectrum $\P_\zeta$, the expression for the abundance of GWs at present time is then simply given by
\begin{align}
h^2 \Omega_\text{\tiny GW}^\TT (\eta_0,k)= c_g \frac{h^2 \Omega_{r,0}}{972} 
 \iint _{\cal S} \d x \d y\, \frac{x^2}{y^2} \, 
\left[ 1 -  \frac{\left(1 + x^2 - y^2  \right)^2}{4 x^2}  \right]^2 
 {\cal P}_\zeta \left( k \, x \right) 
{\cal P}_\zeta \left( k \, y \right)  \, \left[ {{\cal I}^\TT_c}^2 \left( x , y \right) + {{\cal I}^\TT_s}^2 \left( x ,y \right) \right],
\label{eq: Omega GW with PS}
\end{align}
where the domain of integration $\mathcal{S}$ is defined in  \cite{sm}, $c_g \simeq 0.4$ accounts for the change in number of relativistic degrees of freedom and  $\Omega_{r,0} \simeq 5.4 \cdot10^{-5}$ is the current  energy density of the relativistic d.o.f.s in terms of the critical energy density.
For a power spectrum of the curvature perturbation with a Dirac delta shape
\be\label{dirac}
\mathcal{P}_{\zeta} (k)= A k_* \delta \lp  k - k_*\rp,
\ee
the GW abundance at present time is then given by
\begin{align}
\label{omegaGI}
h^2\Omega_\text{\tiny \GW}^\TT (\eta_0,k)= c_g \frac{ h^2\Omega_{r,0}}{15552} A^2  \frac{k^2}{k_*^2}  \lp \frac{4k_*^2}{k^2} -1  \rp^2 
 \left[ {{\cal I}^\TT_c}^2 \left( \frac{k_*}{k} , \frac{k_*}{k} \right) + {{\cal I}^\TT_s}^2 \left(\frac{k_*}{k} ,\frac{k_*}{k} \right) \right]
 \theta (2k_*-k).
\end{align}
In Fig.~\ref{fig: Omega GW} we show that the  gauge invariant power spectrum constructed from the Poisson gauge (which coincides with the standard computation of the power spectrum in the Poisson gauge) and the power spectrum of tensor modes computed in the TT gauge, for two different cases of  curvature perturbation power spectrum \eqref{dirac} and 
\begin{equation}
{\cal P}_\zeta (k) = \frac{A_s}{\sqrt{2 \pi \sigma^2}} {\rm e} ^{-{\rm log}^2(k/k_*)/2 \sigma^2},
\end{equation}
are indeed equal.  This is expected because, as mentioned in the introduction, once the GWs are generated and propagate freely inside our horizon, they can be treated as linear objects and all the gauges must provide the same result. In Eq.~\eqref{omegaGI} this point is made manifest by the fact that the combination $\Ic^2 + \Is^2$ is the same in all gauges. 
The unique results are superimposed with the LISA sensitivity curve which is computed in the TT frame, as we argued in Sec.~\ref{secm}. 


\subsection{Propagation equation in the TT gauge}
The scalar-tensor source in the TT gauge  is 
\begin{align}
s_{ij, \text{\tiny TT}}^{(st)} 
=	-\frac{1}{ a^2(\eta)}
\Big[ &
- \psi_{1} h_{1ij,kk} 
+ 2 h_{1ij} \H   \psi_{1} '
+ \frac{1}{2}   h_{1ij} '  \psi_{1} '
+ 2 h^{k}_{1~(i} \psi_{1,j)k} 
+ h_{1ij}  \psi_{1} ''
- h_{1ij} \psi_{1,kk} 
\nonumber \\
&
+ \frac{1}{2}   h_{1ij} ' \sigma_{1,kk} 
+\sigma_{1,k}  \lp
   h_{1ij,k} ' 
-    h_{1k(i,j)} ' 
 \rp
+2 \psi_{1,k} \lp h_{1k(i,j)} 
-  h_{1ij,k} \rp 
-    h_{1k(i} ' \sigma_{1,j)k}  
\Big]
\end{align}
and, at leading order in derivatives of the tensor field (equivalent to geometrical optics approximation), the propagation equation of motion becomes 
%
%
\begin{tcolorbox}[colframe=white,arc=0pt]
	\vspace{-.4cm}
	\begin{align}
	h^{\TT\prime\prime}_{\lambda} (\eta, \bm k) +  2\H  h^{\TT\prime}_{\lambda} (\eta, \bm k) +k^2  h^{\TT}_{\lambda} (\eta, \bm k)
=
 4 & e_{\lambda}^{ij}(\bm k) \Big[ 
		\psi_1^{\syn}  h_{1ij,kk} ^{\syn}
		- \sigma^{\syn}_{1,k}\lp  h_{1ij,k}^{\syn \prime}- h_{1k(i,j)}^{\syn \prime}\rp
		\Big].
	\end{align}
\end{tcolorbox}
\noindent
As argued above, this is the equation that one should solve in order to find the propagated GWs which are observed at the present ground-based and space-based observatories.
Obviously,  the Shapiro time delay phase picked up during the propagation in a perturbed universe  does not affect the power spectrum \cite{Bartolo:2018rku}. Therefore, even in a perturbed universe, the GW abundance is independent from the gauge.

\section{Conclusions}
\noindent
The issue of the gauge invariance of GWs produced in the early universe arises as soon as such perturbations are generated at second-order in perturbation theory. In this paper we have addressed  this topic by dividing the discussion in three parts. First we have elaborated about the measurement of the GWs and what is the best gauge in which
to calculate the response of the detector to tensor modes. Following Ref. \cite{book} we have argued that the best choice is the so-called TT frame.  
We have pointed out that there is not a unique way to render the GWs gauge invariant; to give an example, we constructed such gauge invariant combination starting from the Poisson gauge, while this is notoriously  not possible in the TT gauge.
 Motivated by the discussion about the measurement, we have also performed the computation of the abundance of GWs in the TT and in the Poisson gauges, showing that they are equal as expected from general grounds. 
 This is the manifestation of the fact that, if the emission takes place in a radiation-dominated universe as in the case considered in the text, then the source rapidly decays and the gravitational waves become freely propagating linear perturbations of the metric. In such a case the gauge dependence vanishes and the physical observable can be easily extracted from the linear tensor modes. Nevertheless, in the different case in which the emission takes place in a matter-dominated era, further studies are needed in order to understand the nature of the observable gravitational signal.

 The topic of the gauge invariance of the gravitational waves at second-order in perturbation theory has been analysed recently also in Refs. \cite{Inomata:2019yww, Yuan:2019fwv}. There the authors computed the GWs abundance in several gauges, and the conclusion reached in this draft agrees with theirs, where the overlap is possible. Ref. \cite{Inomata:2019yww} contains also a detailed discussion on the transfer functions entering the computation in the TT gauge.

Another work will be devoted to the extension at second-order  of the measurement of the GWs \cite{inprep}.

\hfill

\begin{center}
{\bf  Acknowledgments}
\end{center}
\noindent
We thank Michele Maggiore and Sabino Matarrese for useful discussions.
We also thank J. O. Gong for discussions about Ref.~\cite{Gong:2019mui}.
	V.DL., G.F. and A.R. are  supported by the Swiss National Science Foundation (SNSF), project {\sl The non-Gaussian Universe and Cosmological Symmetries}, project number: 200020-178787. The work of A.K. is partially supported by the 
	EDEIL-NTUA/67108600 project. 
	
	\appendix
	
	\section{Metric perturbations at second-order and gauge transformations}\label{appA}
	\noindent
	In this appendix we review and clarify the notation used throughout the paper. We adopt, as a reference, the notation used in \cite{Malik:2008im}.
	
	Throughout the paper we adopt the mostly plus sign notation for the spacetime metric signatures. Furthermore we express the ordinary and covariant derivatives as
	\begin{equation}
		\partial_\nu {\cal O}_\mu \equiv {\cal O}_{\mu,\nu},
		\qquad
	D_\nu {\cal O}_\mu \equiv
\partial_\nu {\cal O}_\mu - \Gamma^\rho_{\mu \nu}{\cal O}_\rho
	\equiv
	 {\cal O}_{\mu;\nu},
	\end{equation}
	respectively. Finally, we use the compact notation for (anti-)symmetrisation with the normalisation coefficient $1/2$ as
	\begin{equation}
		A_{(\mu\nu)} \equiv \frac{1}{2} \lp A_{\mu\nu}+A_{\nu\mu}\rp,
		\qquad
		A_{[\mu\nu]} \equiv \frac{1}{2} \lp A_{\mu\nu}-A_{\nu\mu}\rp.
	\end{equation}
Where not stated otherwise, we indicate with $'$ the ordinary derivative with respect to the conformal time $\eta$.
	
We choose the following convention for the polarisation tensors $e^{ij}_\lambda$ as
\begin{equation}\label{pol}
e^{ij}_L e_{ij}^R = 0 ,
 \quad
e^{ij}_L e_{ij}^L = e^{ij}_R e_{ij}^R = 1
\quad
 \text{and}
 \quad
 h_\lambda(t,\bm{k}) = e_\lambda^{ij}(\bm{k}) \int \d^3x e^{-i\bm{k}\cdot\bm{x}} h_{ij}(t,\bm{x}),
\end{equation}
with 
\begin{equation}
e^{ii}_\lambda(\bm{k}) = 0
\qquad
\text{and}
\qquad
k_i e^{ij}_\lambda(\bm{k}) = 0.
\end{equation}
We also  define the transverse traceless projector $\mathcal{T}_{ij}^{lm}$ as
\begin{equation}
\label{Tproj}
\mathcal{T}_{ij}^{lm} = e^L_{ij} \otimes e_L^{lm} + e^R_{ij} \otimes  e_R^{lm}.
\end{equation}

	\subsection{Perturbations of the metric up to second-order}
	\noindent
The background metric $\bar g_{\mu \nu}$ for an FRWL cosmology can be written as
\begin{equation}
	\d s^2 \equiv \bar g_{\mu \nu} \d x^\mu \d x^\nu= - a^2(\eta) \llp \d \eta^2 + \delta_{ij}\, \d x^i \d x^j\rrp,
\end{equation}
where $\eta$ is the conformal time. 
The background evolution is described by the Friedmann equations
\begin{align}
\H^2 = \frac{8 \pi G}{3} a^2 \rho,
\qquad 
\H' = -\frac{4\pi G}{3}  a^2 (\rho + 3 P),
\end{align}
and also one gets using the equation of state $P = w\rho$, 
\begin{equation}
\H' - \H^2 = -4\pi G a^2 (\rho + P)
\qquad \text{and} \qquad
\H'=- \H^2 \lp 1+ 3 w \rp /2.
\end{equation}
The perturbed metric $g_{\mu \nu} \equiv \bar g_{\mu \nu} +\delta g_{\mu \nu}$ can be decomposed in the following quantities
\begin{align}\label{metricpert}
\delta g_{00} = -2 a^2 \phi, \qquad \delta g_{0i} = a^2 B_i, \qquad \delta g_{ij} = 2 a^2 C_{ij},
\end{align}
where, when expanded in the scalar-vector-tensor (SVT) decomposition, one can define
\begin{align}
B_i = B_{,i} - S_i, \qquad C_{ij} = -\psi \delta_{ij} + E_{,ij} + F_{(i,j)} + \frac{1}{2}h_{ij}.
\end{align}
The vector and tensor degrees of freedom are defined transverse (divergence-free) and traceless, which means that they satisfy the conditions
\begin{equation}
	S_{i,i} = 0,
	\qquad
		F_{i,i} = 0,
		\qquad
		\text{and}
		\qquad
	h^i_{i}=	h_{ij,j} = 0.
\end{equation}
One can define also quantities at first and second-order in perturbations around the background as
\begin{subequations}
\begin{align}
	\phi &= \phi_1 + \frac{1}{2} \phi_2 + \dots
	\\
	\psi &= \psi_1 + \frac{1}{2} \psi_2 + \dots
	\\
	B &= B_1 + \frac{1}{2} B_2 + \dots
	\\
	E  &= E_1 + \frac{1}{2} E_2 + \dots
	\\
	S_i  &= S_{1i} + \frac{1}{2} S_{2i} + \dots
	\\
	F_i  &= F_{1i} + \frac{1}{2} F_{2i} + \dots
	\\
	h_{ij} &= h_{1ij} +\frac{1}{2} h_{2ij} + \dots
\end{align}	
\end{subequations}
The connection coefficients, defined as 
\begin{equation}
\Gamma^{\rho}	_{\mu \nu} = \frac{1}{2} g^{\rho \sigma}\lp g_{\mu \sigma, \nu} +g_{\nu \sigma, \mu} -g_{\mu \nu, \sigma}\rp
\end{equation}
can be expressed, up to second-order, as\footnote{We acknowledge the use of the symbolic tensor computation Mathematica package xAct \cite{xact} to perform many of the calculations presented in the paper.}
\begin{align}
\Gamma^0_{00}&=
\H + \epsilon    \phi_1 '+
 \epsilon ^2 (B_{1b} B_1^{b} \H 
 + B_1^{b}   B_{1b}'
  -2 \phi_1   \phi_1'
   + \frac{1}{2}   \phi_2'
    + B_1^{b} \partial _{b}\phi_1) + \mathcal{O}(\epsilon ^3),
\nonumber \\
\Gamma^0_{0i}&=
\epsilon  (B_{1i} \H + \partial _{i}\phi_1) 
+ \frac{1}{2} \epsilon ^2 (B_{2i} \H 
-4 B_{1i} \H \phi_1 + 2 B_1^{b}  C_{1ib} '
- B_1^{b} \partial _{b}B_{1i} + 
B_1^{b} \partial _{i}B_{1b} 
-4 \phi_1 \partial _{i}\phi_1 
+ \partial _{i}\phi_2) + \mathcal{O}(\epsilon ^3),
\nonumber \\
\Gamma^i_{00}&=
\epsilon  (B_1^{i} \H +   B_1^{i \prime} + 
\partial^{i}\phi_1) 
+ \frac{1}{2} \epsilon ^2 (B_2^{i} \H 
- 4 B_1^{b} C_{1b}{}^{i} \H 
 -4 C_1^{ib}  B_{1b}' +
    B_2^{i\prime} -2 B_1^{i}   \phi_1  '
-4 C_1^{ib} \partial _{b}\
\phi_1 + \partial ^{i}\phi_2) + \mathcal{O}(\epsilon ^3),
\nonumber \\
\Gamma^0_{ij}&=
g_{ij} \H +
 \epsilon  (2 C_{1ij}  \H -2 g_{ij} \H \phi_1 
 +  C_{1ij}'  - \frac{1}{2} 
\partial _{i}B_{1j} - \frac{1}{2} \partial _{j}B_{1i}) 
+ \epsilon ^2 (C_{2ij}  \H - B_{1b} B_1^{b} g_{ij} \H -4 C_{1ij}  \H \phi_1
\nonumber \\
&
+ 4 g_{ij} \H \phi_1^2 - g_{ij} \H \phi_2 -2 \phi_1  C_{1ij} ' + \frac{1}{2}  C_{2ij} ' - B_1^{b} \partial_{b}C_{1ij}  + \phi_1 \partial _{i}B_{1j} - \frac{1}{4} \partial_{i}B_{2j} + B_1^{b} \partial _{i}C_{1jb}
\nonumber \\
&
+ \phi_1 \partial_{j}B_{1i} - \frac{1}{4} \partial _{j}B_{2i} + B_1^{b} \partial_{j}C_{1ib}) + \mathcal{O}(\epsilon ^3),
\nonumber \\
\Gamma^i_{0j}&=
\delta ^{i}{}_{j} \H +
 \epsilon  (
 C_{1j}{}^{i\prime} - \frac{1}{2} \partial ^{i}B_{1j} + \frac{1}{2} \partial _{j}B_1^{i}) + \frac{1}{4} \epsilon ^2 (-4 B_1^{i} B_{1j} \H -8 C_1^{ib}  C_{1jb}' + 2 C_{2j}{}^{i\prime } + 4 C_1^{ib} \partial _{b}B_{1j} 
\nonumber \\
&- \partial ^{i}B_{2j} -4 C_1^{ib} \partial _{j}B_{1b} + \partial _{j}B_2^{i} -4 B_1^{i} \partial _{j}\phi_1) + \mathcal{O}(\epsilon ^3),
\nonumber \\
\Gamma^i_{jk}&=\epsilon  (- B_1^{i} g_{jk} \H - \partial ^{i}C_{1jk} + \partial _{j}C_{1k}{}^{i} + \partial _{k}C_{1j}{}^{i}) + \frac{1}{2} \epsilon ^2 (-4 B_1^{i} C_{1jk} \H - B_2^{i} g_{jk} \H + 4 B_1^{b} C_{1b}{}^{i} g_{jk} \H
\nonumber \\
&
 + 4 B_1^{i} g_{jk} \H \phi_1 -2 B_1^{i}   C_{1jk} '
 + 4 C_1^{ib} \partial _{b}C_{1jk} - \partial ^{i}C_{2jk} + B_1^{i} \partial _{j}B_{1k} -4 C_1^{ib} \partial _{j}C_{1kb} + \partial _{j}C_{2k}{}^{i} 
\nonumber \\
&
+ B_1^{i} \partial _{k}B_{1j} -4 C_1^{ib} \partial _{k}C_{1jb} + \partial _{k}C_{2j}{}^{i}) + \mathcal{O}(\epsilon ^3),
\end{align}
where, for clarity, we introduced the bookeeping parameter $\epsilon$ specifying each order in perturbation theory. The mixed components $\{i0j0\}$ of the Riemann tensor defined as
\be
R^{\alpha}_{~\beta\mu\nu}=
\Gamma^{\alpha}_{\beta\nu,\mu}-\Gamma^{\alpha}_{\beta\mu,\nu}+
\Gamma^{\alpha}_{\lambda\mu}\Gamma^{\lambda}_{\beta\nu}-
\Gamma^{\alpha}_{\lambda\nu}\Gamma^{\lambda}_{\beta\mu} ,
\ee
are given, up to second-order, by
\begin{align}
R_{i0j0}=&
- a^2 g_{ij} \partial_\tau  \mathcal{H}  
\nonumber \\
&+ \frac{1}{4} \epsilon  a^2 \llp
-4 \mathcal{H}    C_{1ij} '
 -8 C_{1ij}  \mathcal{H} '
  + 4 g_{ij} \mathcal{H}   \phi_1 '
  + 2 \partial _{i}B_{1j} '
   +2   \partial _{j}B_{1i} '
    -4 C_{1ij} ''
    + 2 \mathcal{H}  \partial_{i}B_{1j} 
    + 2 \mathcal{H}  \partial _{j}B_{1i} 
     + 4 \partial _{j}\partial _{i}\phi_1
    \rrp
\nonumber \\ &
+ \frac{1}{4} \epsilon ^2 a^2 
\Big[ 4 B_{1b} B_1^{b} g_{ij} \mathcal{H} ^2 
+4 B_1^{b} g_{ij} \mathcal{H}    B_{1b} '
+ 4  C_{1i}{}^{b\prime}   C_{1jb}' 
-2 \mathcal{H}    C_{2ij}' 
 -4 C_{2ij}  \mathcal{H} ' 
 + 8 C_{1ij} \mathcal{H}   \phi_1' 
  -8 g_{ij} \mathcal{H}  \phi_1   \phi_1' 
  \nonumber \\ &
    + 4   C_{1ij}'   \phi_1' 
  + 2 g_{ij} \mathcal{H}    \phi_2'
   + \partial _{i}B_{2j}' 
    +  \partial _{j}B_{2i} ' 
     -2 C_{2ij} '' 
     + 4 B_1^{b} \mathcal{H}  \partial_{b}C_{1ij} 
     + 4  B_1^{b \prime}\partial _{b}C_{1ij} 
     + 4 B_1^{b} g_{ij} \mathcal{H}  \partial _{b}\phi_1
       -2  C_{1jb}' \partial ^{b}B_{1i} 
\nonumber \\ &
 + \partial _{b}B_{1j} \partial ^{b}B_{1i} 
        -2   C_{1ib}' \partial ^{b}B_{1j} 
        + 4 \partial _{b}C_{1ij} \partial ^{b}\phi_1 
        + 2   C_{1jb}' \partial _{i}B_1^{b} 
         - \partial _{b}B_{1j} \partial _{i}B_1^{b} 
-2   \phi_1 '\partial _{i}B_{1j} 
+ \mathcal{H}  \partial _{i}B_{2j} 
 -4 B_1^{b} \mathcal{H}  \partial _{i}C_{1jb} 
 \nonumber \\ &
  -4  B_1^{b \prime} \partial _{i}C_{1jb} 
   -4 \partial ^{b}\phi_1 \partial _{i}C_{1jb} 
    -4 B_{1j} \mathcal{H}  \partial _{i}\phi_1
     - \partial ^{b}B_{1i} \partial _{j}B_{1b} 
     + \partial _{i}B_1^{b} \partial _{j}B_{1b} 
     + 2   C_{1ib} ' \partial _{j}B_1^{b} 
      -2   \phi_1 ' \partial _{j}B_{1i} 
      + \mathcal{H}  \partial _{j}B_{2i} 
      \nonumber \\ &
       -4 B_1^{b} \mathcal{H}  \partial _{j}C_{1ib}  
       -4   B_1^{b \prime} \partial _{j}C_{1ib}  
        -4 \partial ^{b}\phi_1 \partial _{j}C_{1ib}  
         -4 \partial _{i}\phi_1 \partial _{j}\phi_1 
-4 B_{1i} \mathcal{H}  (B_{1j} \mathcal{H}  
+ \partial _{j}\phi_1) 
+ 2 \partial _{j}\partial _{i}\phi_2
\Big ].
\end{align}

\subsection{Gauge transformations up to second-order}\label{app2}
\noindent
One can perform a gauge transformation of the form $x^\mu \to \widetilde x^\mu =x^\mu + \xi^\mu$ with 
\begin{equation}
	\xi^\mu = \lp \alpha_1+ \alpha_2 , \xi^i_1 + \xi^i_2\rp 
	\qquad 
	\text{and}
	\qquad 
	\xi^i_a= \beta_{a}^{i} + \gamma_{a}^{i},
\end{equation}
where $a=\{1,2\}$ and  where we defined a  divergence-free vectorial parameter such that $\gamma^i_{i}=0$.

The first-order gauge transformations are given by \cite{Malik:2008im}
\begin{subequations}
\label{gt1o}
\begin{align}
\label{transphi1}
\widetilde {\A1} =& \A1 +\H\alpha_1+\alpha_1'\,,\\
\label{transpsi1}
\widetilde \psi_1 =& \psi_1-\H\alpha_1 \,,\\
\label{transB_1}
\widetilde B_1 =& B_1-\alpha_1+\beta_1'\,,\\
\label{transE_1}
\widetilde E_1 =& E_1+\beta_1\,,\\
\label{transS1}
\widetilde {S_{1}^{~i}} =& S_{1}^{~i}-\gami1'\,, \\
\label{transF1}
\widetilde {F_{1}^{~i}} =& F_{1}^{~i}+\gami1\,,\\
\widetilde h_{1ij} =& h_{1ij}\,.
\end{align}
\end{subequations}
At second-order the gauge transformation can be written by defining 
the vector $\XB_i$ and tensor $\X_{ij}$ (dependent only on first-order quantities squared) as
\begin{align}
\label{defXBi}
\XB_i
&\equiv 
2\Big[
\left(2\H B_{1i}+B_{1i}'\right)\alpha_1
+B_{1i,k}\xi_1^k-2\phi_1\alpha_{1,i}+B_{1k}\xi_{1,~i}^k
+B_{1i}\alpha_1'+2 C_{1ik}{\xi_{1}^k}'
 \Big]\nonumber\\
&+4\H\alpha_1\left(\xi_{1i}'-\alpha_{1,i}\right)
+\alpha_1'\left(\xi_{1i}'-3\alpha_{1,i}\right)
+\alpha_1\left(\xi_{1i}''-\alpha_{1,i}'\right)\nonumber\\
&+{\xi_{1}^k}'\left(\xi_{1i,k}+2\xi_{1k,i}\right)
+\xi_{1}^k\left(\xi_{1i,k}'-\alpha_{1,ik}\right)
-\alpha_{1,k}\xi_{1,i}^k
\end{align}
and 
\begin{align}
\label{Xijdef}
\X_{ij}&\equiv
2\Big[\left(\H^2+\frac{a''}{a}\right)\alpha_1^2
+\H\left(\alpha_1\alpha_1'+\alpha_{1,k}\xi_{1}^{~k}
\right)\Big] \delta_{ij}\nonumber\\
&
+4\Big[\alpha_1\left(C_{1ij}'+2\H C_{1ij}\right)
+C_{1ij,k}\xi_{1}^{~k}+C_{1ik}\xi_{1~~,j}^{~k}
+C_{1kj}\xi_{1~~,i}^{~k}\Big]
+2\left(B_{1i}\alpha_{1,j}+B_{1j}\alpha_{1,i}\right)
\nonumber\\
&
+4\H\alpha_1\left( \xi_{1i,j}+\xi_{1j,i}\right)
-2\alpha_{1,i}\alpha_{1,j}+2\xi_{1k,i}\xi_{1~~,j}^{~k}
+\alpha_1\left( \xi_{1i,j}'+\xi_{1j,i}' \right)
+\left(\xi_{1i,jk}+\xi_{1j,ik}\right)\xi_{1}^{~k}
\nonumber\\
&+\xi_{1i,k}\xi_{1~~,j}^{~k}+\xi_{1j,k}\xi_{1~~,i}^{~k}
+\xi_{1i}'\alpha_{1,j}+\xi_{1j}'\alpha_{1,i},
\end{align}
such that one finds
\begin{subequations}
\begin{align}
\label{transphi2}
\wt {\A2} &= \A2+\H\alpha_2+{\alpha_2}'
+\alpha_1\left[{\alpha_1}''+5\H{\alpha_1}' +\left(\H'+2\H^2
\right)\alpha_1 +4\H\phi_1+2\phi_1'\right]\nonumber \\
&+2{\alpha_1}'\left({\alpha_1}'+2\phi_1\right)
+\xi_{1k}
\left({\alpha_1}'+\H{\alpha_1}+2\phi_1\right)_{,}^{~k}
+\xi_{1k}'\left[\alpha_{1,}^{~k}-2B_{1k}-{\xi_1^k}'\right], \\
\label{transpsi2}
\wt\psi_2&=\psi_2-\H\alpha_2-\frac{1}{4}\X^k_{~k}
+\frac{1}{4}\nabla^{-2} \X^{ij}_{~~,ij},
\\
\label{transB2}
\widetilde B_{2} &=B_{2}-\alpha_2+\beta_2' +\nabla^{-2} \XB^k_{~,k},
\\
\label{transE_2}
\wt E_2&=E_2+\beta_2+\frac{3}{4}\nabla^{-2}\nabla^{-2}\X^{ij}_{~~,ij}
-\frac{1}{4}\nabla^{-2}\X^k_{~k},
\\
\label{transS2}
\widetilde S_{2i}&=S_{2i}-\gamma_{2i}'-\XB_i+\nabla^{-2}\XB^k_{~,ki},
\\
\label{transFi2}
\wt F_{2i}&= F_{2i}+\gamma_{2i}
+\nabla^{-2}\X_{ik,}^{~~~k}-\nabla^{-2}\nabla^{-2}\X^{kl}_{~~,kli},
\\
\label{transhij2}
\wt h_{2ij}&= h_{2ij}+\X_{ij}
+\frac{1}{2}\left(\nabla^{-2}\X^{kl}_{~~,kl}-\X^k_{~k}
\right)\delta_{ij}
+\frac{1}{2}\nabla^{-2}\nabla^{-2}\X^{kl}_{~~,klij}
+\frac{1}{2}\nabla^{-2}\X^k_{~k,ij}
-\nabla^{-2}\left(\X_{ik,~~~j}^{~~~k}+\X_{jk,~~~i}^{~~~k}
\right).
\end{align}
\end{subequations}

\subsection{Explicit expression for $\X^\PP_{ij}$}
Here, for completeness, we report the explicit field combinations for  $\X_{ij}^\PP$ as
\begin{align}
\label{XPP}
\X_{ij}^\PP = &
4 h_{1ij} \H  (B_{1} -   E_{1} ') 
+ 2 B_{1}  h_{1ij} '
-2  E_{1} '   h_{1ij}' 
+ 2 B_{1}   F_{1j,i} '
-2   E_{1}'   F_{1j,i} '
+ 2 B_{1}   F_{1i,j} '
-2   E_{1}'   F_{1i,j} '
+ 4 B_{1}  E_{1,ij} '
-2   E_{1} '  E_{1,ij} '
\nn \\ &
-2 S1_{j} B_{1,i} 
+   {{\cal S}_1}_{j}' B_{1,i} 
-   E_{1,j} ' B_{1,i} 
+ 4 B_{1} \H  F_{1j,i} 
-4 \H  E_{1}' F_{1j,i} 
+ 4 B_{1} \H {{\cal S}_1}_{j,i} 
-4 \psi_{1} {{\cal S}_1}_{j,i} 
-4 \H  E_{1}' {{\cal S}_1}_{j,i} 
+ 2 h_{1jk} {{\cal S}_1}^{k}_{,i}
\nn \\ &
+ 2 S1_{j} E_{1,i}' 
- {{\cal S}_1}_{j}' E_{1,i}' 
+ B_{1} {{\cal S}_1}_{j,i}' 
- E_{1}' {{\cal S}_1}_{j,i}' 
-2  B_{1} E_{1,ij}' 
-2 S_{1i} B_{1,j} 
+ {{\cal S}_1}_{i}' B_{1,j} 
- E_{1,i}' B_{1,j} 
+ 2 B_{1,i} B_{1,j} 
+ 4 B_{1} \H  F_{1i,j} 
\nn \\ &
-4 \H  E_{1}' F_{1i,j} 
+ 2 {{\cal S}_1}_{k,i} F_{1,j}^{k} 
+ 4 B_{1} \H  {{\cal S}_1}_{i,j} 
-4 \psi_{1} {{\cal S}_1}_{i,j} 
-4 \H  E_{1}' {{\cal S}_1}_{i,j} 
+ 2 F_{1,i}^{k} {{\cal S}_1}_{k,j} 
+ 2 {{\cal S}_1}^{k}_{,i}{{\cal S}_1}_{k,j} 
+ 2 h_{1ik} {{\cal S}_1}^{k}_{,j} 
+ 2 S_{1i} E_{1,j}' 
\nn \\ &
- {{\cal S}_1}_{i}' E_{1,j}' 
+ B_{1} {{\cal S}_1}_{i,j}' 
- E_{1}' {{\cal S}_1}_{i,j}' 
+ 8 \psi_{1} E_{1,ij} 
+ 2 {{\cal S}_1}^{k} h_{1ij,k} 
-2 F_{1,j}^{k} E_{1,ik} 
+ {{\cal S}_1}^{k}_{,j} E_{1,ik} 
+ 2 {{\cal S}_1}^{k} F_{1j,ik} 
+ {{\cal S}_1}^{k} {{\cal S}_1}_{j,ik} 
\nn \\ &
-2 F_{1,i}^{k} E_{1,jk} 
+ {{\cal S}_1}^{k}_{,i}E_{1,jk}  
+ 2 {{\cal S}_1}^{k} F_{1i,jk}
+ {{\cal S}_1}^{k} {{\cal S}_1}_{i,jk}
 + 2 {{\cal S}_1}^{k} E_{1,ijk} 
-2 h_{1ij,k} E_{1,k} 
-2 F_{1j,ik} E_{1,k} 
- {{\cal S}_1}_{j,ik} E_{1,k} 
\nn \\ &
-2 F_{1i,jk} E_{1,k} 
- {{\cal S}_1}_{i,jk} E_{1,k} 
-2 E_{1,ijk} E_{1,k} 
+ 2 g_{ij} \big [2 B_{1}^2 \H ^2 
-4 B_{1} \H  \psi_{1} 
+ \H  B_{1}' (B_{1} - E_{1}') 
-4 B_{1} \H ^2 E_{1}' 
+ 4 \H  \psi_{1} E_{1}' 
\nn \\ &
+ B_{1}^2 \H' 
-2 B_{1} E_{1}' \H'  
+ (E_{1}')^2 (2 \H ^2 + \H' ) 
-2 B_{1} \psi_{1}' 
+ 2 E_{1}' \psi_{1}' 
- B_{1} \H  E_{1}'' 
+ \H  E_{1}' E_{1}'' 
+ {{\cal S}_1}^{k} \H  B_{1,k}  
-2 {{\cal S}_1}^{k} \psi_{1,k} 
\nn \\ &
- {{\cal S}_1}^{k} \H  E_{1,k}' 
- \H  E_{1,k} B_{1,k} 
+ 2 \psi_{1,k} E_{1,k} 
+ \H  E_{1,k}' E_{1,k}
\big]
+ 2 {{\cal S}_1}_{k,j} F_{1i,k}  
-2 E_{1,jk} F_{1i,k}  
+ 2 {{\cal S}_1}_{k,i} F_{1j,k} 
-2 E_{1,ik} F_{1j,k} 
\nn \\ &
+ {{\cal S}_1}_{k,j} {{\cal S}_1}_{i,k} 
- E_{1,jk} {{\cal S}_1}_{i,k} 
+ {{\cal S}_1}_{k,i} {{\cal S}_1}_{j,k} 
- E_{1,ik} {{\cal S}_1}_{j,k} 
-2 h_{1jk} E_{1,ik} 
-4 E_{1,jk} E_{1,ik} 
-2 h_{1ik} E_{1,jk}
\end{align}
where we defined ${{\cal S}_1}_{i} \equiv \int^\eta  S_{1i} \d\eta' + \hat\C_{1i}(\bm{x})$.

\section{First-order dynamics in the Poisson gauge}
\noindent
\label{app3}
In this appendix we collect the equations of motion for the perturbations in the Poisson gauge at first-order. 
\begin{itemize}
	\item {\bf Scalar perturbations:} 
 the perturbed Einstein equations at first-order provide two
evolution equations for the scalar metric perturbations \cite{Malik:2008im}
\begin{align}
 \label{eq:psievol}
\psi_1''+2\H\psi_1'+\H\A{}_1'
+ \left( 2\H'+\H^2 \right) \A{}_1
&=4\pi G a^2 \left(\delta P+\frac{2}{3}\nabla^2\Pi_1\right),\\
\label{eq:aniso}
\sigma_1'+2\H \sigma_1+\psi_1-\A{}_1&=8\pi G a^2 \Pi_1,
\end{align}
where $\Pi_1$ is the scalar part of the (trace-free) first-order anisotropic stress.
The latter equation can be written in terms of the gauge invariant Bardeen's potentials (equivalent to $\phi_1$ and $\psi_1$ in the Poisson gauge) as  
\begin{equation}
\Psi_1 - \Phi_1 =
8\pi G a^2 \Pi_1,
\end{equation}
which implies that  $\Psi_1=\Phi_1$ in the absence of anisotropic stress. 
For the case of adiabatic perturbations, for which one has $\delta P=c_s^2\delta\rho$, one can
insert this result into the former equation and, neglecting the anisotropic stress and by using the energy momentum constraints, one finds 
\begin{equation}
\Psi_1''+3(1+c_s^2)\H\Psi_1' + [2\H'+(1+3c_s^2)\H^2-c_s^2\nabla^2]\Psi_1 =
0.
\end{equation}
In the Poisson gauge, the solution of the previous equation in radiation-dominated is given by the linear transfer function expressed in terms of the comoving curvature perturbation $\zeta$ 
\begin{eqnarray}
\label{transfer}
\Psi_1(\eta,k) \equiv  \frac{2}{3} \zeta(k) T(\eta k)
=\frac{2}{3} \zeta(k) 3\frac{j_1(z)}{z}
=
 \frac{2}{3} \zeta(k) 3\frac{\sin \lp  k \eta/\sqrt{3} \rp - \lp  k \eta/\sqrt{3} \rp \cos \lp  k \eta/\sqrt{3} \rp}{\lp  k \eta/\sqrt{3} \rp^3},
\end{eqnarray}
where $j_1(z)$ is the spherical Bessel function and $z =  \eta k/\sqrt{3}$.

 The equation of motion for the velocity potential  $v$ in the Poisson gauge is written as
\begin{equation}
\Psi_1' + \H \Psi_1 = -4\pi G a^2 (\rho + P)v.
\end{equation}

\item {\bf Vector perturbations:} the gauge invariant vector
metric perturbation is directly related to the divergence-free
part of the momentum via the constraint equation
\begin{equation}
 \label{eq:convec}
\nabla^2 \left( {F}_{1i}' + S_{1i} \right) = - 16\pi G a^2 \delta q_i,
\end{equation}
where the momentum conservation equation yields
\begin{equation}
\label{eq:evolvec}
{\delta q}_i' + 4\H \delta q_i = - \nabla^2 \Pi_{1i}.
\end{equation}
Therefore, in the absence of anisotropic stress sourcing ${\delta q}_i$, the vector perturbations are rapidly redshifted with the Hubble expansion as
\begin{equation}
{\delta q}_i (\eta)={\delta q}_i (\eta_\text{\tiny in}) /a^4(\eta).
\end{equation}
In particular, in both Poisson and TT gauge where one sets $S_{1i} = 0$, one reads directly from \eqref{eq:convec} that the vector perturbations left are redshifted away in an expanding universe.

\item {\bf Tensor perturbations:}  from the linearised Einstein's equations one finds
\begin{equation}
\label{teneq}
{h}_{1ij}'' + 2\H{h}_{1ij}' - {\nabla^2} h_{1ij} = 8\pi G a^2
\Pi_{1ij} \,,
\end{equation}
which becomes the standard homogeneous evolution equation for the linear tensor modes in the absence of the anisotropic stress.
\end{itemize}

\section{Transfer functions for the GW abundance in the TT gauge}
\label{appc}
\begin{figure}[t!]
	\centering
	\includegraphics[width=0.48\columnwidth]{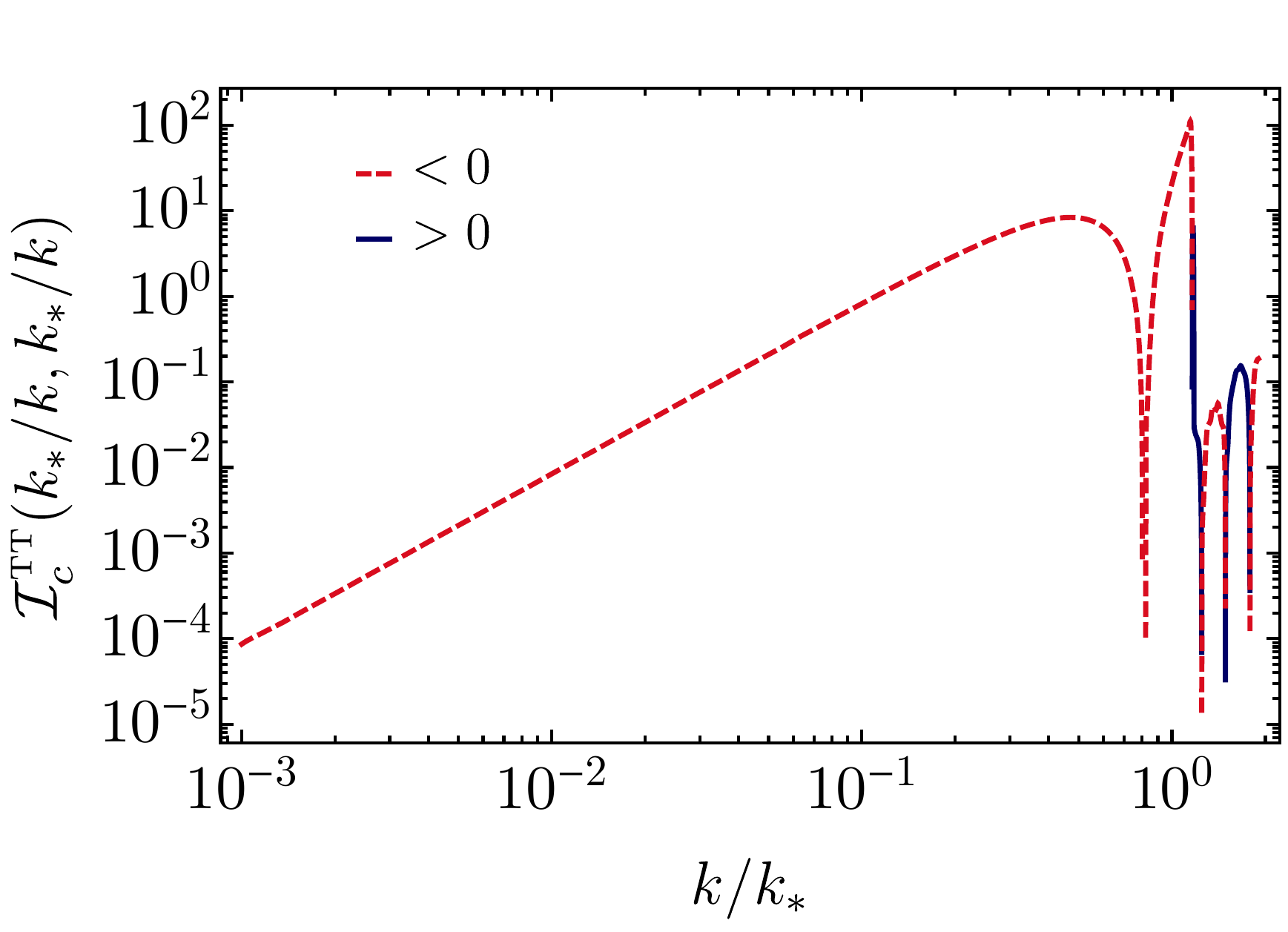}
	\hspace{0.3 cm}
	\includegraphics[width=0.48\columnwidth]{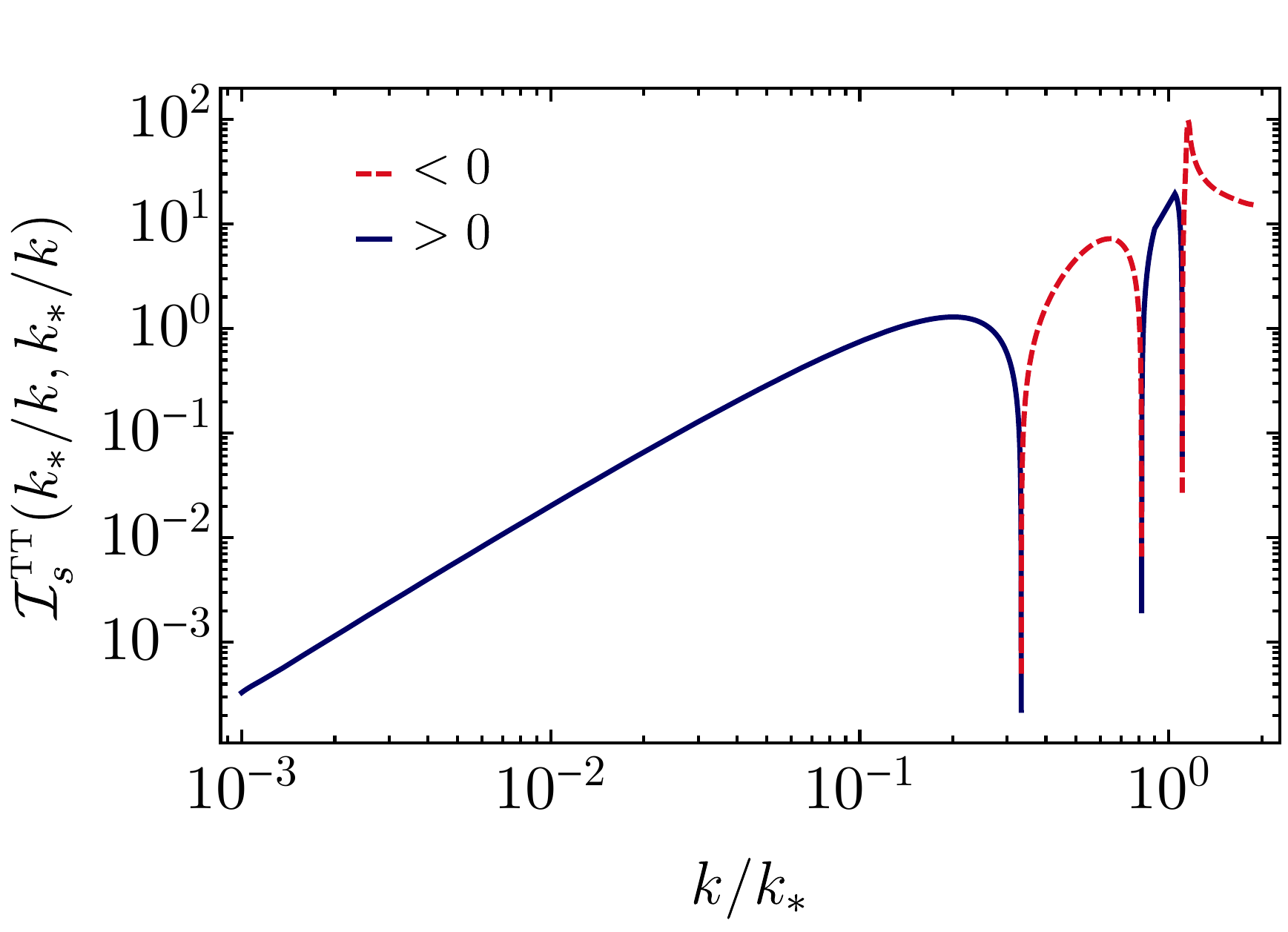}
	\caption{The oscillating functions ${\cal I}^\TT_c$ and ${\cal I}^\TT_s$ for the choice of the arguments $x = y = k_*/k$ (relevant for the case of a Dirac delta power spectrum of the curvature perturbation).
	}
	\label{figIcIs}
\end{figure}

\noindent
In this appendix we provide some details about the transfer functions which enter the calculation of the GWs abundance in the TT gauge in a radiation-dominated universe.
Given the expression for the transfer functions in the TT gauge, which we report here in terms of $z = k \eta/\sqrt{3}$,
\begin{align}
T_\psi(z) & = 3\llp \frac{j_1(z)}{z} - \frac{j_0(z)}{z^2}
+ \frac{1 }{z^2} \rrp, \nonumber \\
T_\sigma(z) & =
3 \llp \frac{j_0(z)}{z} -\frac{1}{z} \rrp,
\end{align}
the function in Eq. \eqref{fTT} is then explicitly given by 
\begin{align}
f^{\TT}(x,y,u) & = \frac{54}{u^5 x^3 y^2} \left\{3 \left[u x \left(u^2 \left(x^2-y^2-1\right)-2\right)+\sqrt{3} \left(u^2
	\left(-x^2+y^2+1\right)+4\right) \sin \left(\frac{u x}{\sqrt{3}}\right)-2 u x \cos
	\left(\frac{u x}{\sqrt{3}}\right)\right] \right. \nonumber \\
	& \left. +\frac{1}{y}\sin \left(\frac{u y}{\sqrt{3}}\right)
		\left[\sqrt{3} x \left(u^2 \left(-3 x^2+y^2+3\right)+24\right)+u \left(y^2 \left(2 u^2
		x^2-15\right)-3 \left(x^2+3\right)\right) \sin \left(\frac{u x}{\sqrt{3}}\right) \right. \right. \nonumber \\
	& \left. \left.	+4\sqrt{3} x \left(u^2 y^2-6\right) \cos \left(\frac{u x}{\sqrt{3}}\right)\right]+2
	\cos \left(\frac{u y}{\sqrt{3}}\right) \left[2 \sqrt{3} \left(u^2 x^2-3\right) \sin
	\left(\frac{u x}{\sqrt{3}}\right)-9 u x+15 u x \cos \left(\frac{u
		x}{\sqrt{3}}\right)\right]\right\}
\end{align}
and the corresponding oscillating functions ${\cal I}^\TT_c$ and ${\cal I}^\TT_s$, defined as
\begin{align}
\mathcal{I}^{\TT}_c(x,y) & = \int_0^\infty \d u \, u (-\sin u) f^{\TT}(x,y,u), \nonumber \\
\mathcal{I}^{\TT}_s(x,y) & = \int_0^\infty \d u \, u (\cos u) f^{\TT}(x,y,u),
\end{align}
 are plotted in Fig. \ref{figIcIs}.
\noindent



\end{document}